\documentclass[preprint]{aastex}
\newcommand{\kms}{\mbox{km~s$^{-1}$}}
\newcommand{\Msun}{\mbox{$\,M_{\sun}$}}
\newcommand{\Lsun}{\mbox{$\,L_{\sun}$}}

\newcommand{\cs}[2]{\mbox{CS~#1$\rightarrow$#2}}
\newcommand{\hii}{H\,{\rmfamily\scshape{ii}}}
\newcommand{\ha}[1]{H#1$\alpha$}
\newcommand{\hb}[1]{H#1$\beta$}
\newcommand{\Tau}{\mathcal{T}}

\newcommand{\chg}[1]{{#1}}

\shortauthors{Guzm{\'a}n et al.}
\shorttitle{Slow ionized wind and disk towards G345.49+1.47}
\usepackage{amssymb,amsmath,natbib,graphicx}
\begin{document}

%%%%%%%%%%%%%%%%%%%%%%%%%%%%%%%%%%%%%%%%%%%%%%%%%%%%%%%%%%%%%%%%%%%%%%%%%
% TITLE and AUTHORS
%%%%%%%%%%%%%%%%%%%%%%%%%%%%%%%%%%%%%%%%%%%%%%%%%%%%%%%%%%%%%%%%%%%%%%%%%

\title{Slow ionized wind and rotating disklike system associated with the high-mass young stellar object G345.4938+01.4677}

\author{Andr\'es E. Guzm{\'{a}}n\altaffilmark{1,2}, Guido
  Garay\altaffilmark{1}, Luis F. Rodr\'iguez\altaffilmark{3}, James
  Moran\altaffilmark{2}, Kate J. Brooks\altaffilmark{4}, Leonardo
  Bronfman\altaffilmark{1}, Lars-{\AA}ke Nyman\altaffilmark{5}, Patricio Sanhueza \altaffilmark{6,7}, and Diego
  Mardones\altaffilmark{1}.}
\altaffiltext{1}{Departamento de Astronom\'{\i}a, Universidad de Chile, Camino el
  Observatorio 1515, Las Condes, Santiago, Chile}%
\altaffiltext{2}{Harvard-Smithsonian Center for Astrophysics, 60 Garden Street, Cambridge, MA, USA}%
\altaffiltext{3}{Centro de Radioastronom\'\i a y Astrof\'\i sica (UNAM), Morelia 58089, M\'exico}%
\altaffiltext{4}{CSIRO Astronomy and Space Science, P.O. Box 76, Epping 1710 NSW,  Australia}%
\altaffiltext{5}{Joint ALMA Observatory (JAO), Alonso de C\'ordova 3107, Vitacura, Santiago, Chile}
\altaffiltext{6}{Institute for Astrophysical Research, Boston University, Boston, MA, USA}
\altaffiltext{7}{National Astronomical Observatory of Japan, 2-21-1 Osawa, Mitaka, Tokyo 181-8588, Japan}
%%%%%%%%%%%%%%%%%%%%%%%%%%%%%%%%%%%%%%%%%%%%%%%%%%%%%%%%%%%%%%%%%%%%%%%%%
% Abstract
%%%%%%%%%%%%%%%%%%%%%%%%%%%%%%%%%%%%%%%%%%%%%%%%%%%%%%%%%%%%%%%%%%%%%%%%%
\begin{abstract}
We report the detection, made using ALMA, of the 92 GHz continuum and
hydrogen recombination lines (HRLs) H40$\alpha$, H42$\alpha$, and
H50$\beta$ emission toward the ionized wind associated with the
high-mass young stellar object G345.4938+01.4677.  This is the
luminous central dominating source located in the massive and dense
molecular clump associated with IRAS 16562$-$3959. The HRLs exhibit
Voigt profiles, a strong signature of Stark broadening.
\chg{We successfully reproduce the observed continuum and HRLs
  simultaneously using a simple model of a slow ionized wind in local
  thermodynamic equilibrium, with no need a high-velocity component.} 
The Lorentzian
line wings imply electron densities of $5\times10^7$~cm$^{-3}$ on
average.  
 In addition, we detect SO and SO$_2$
emission arising from a compact ($\sim3000$~AU) molecular core
associated with the central young star. The molecular core exhibits a
velocity gradient perpendicular to the jet-axis, which we interpret as
evidence of rotation. The set of observations toward G345.4938+01.4677
are consistent with it being a young high-mass star associated with a
slow photo-ionized  wind.
\end{abstract}

\vfill\eject

\section{INTRODUCTION}

Stars of all masses form by gravitational collapse within unstable
regions of molecular clouds. Observationally, low-mass star formation
 is characterized by the following inter-related phenomena: an
infalling envelope, an accretion disk, and a highly collimated jet
\citep{Shu1987ARA&A,Li2014arXiv}.  Highly collimated jets, flowing outwards in a
roughly symmetrical fashion are one of the most spectacular phenomena
occurring during the formation of stars
\citep{DeYoung1991Sci}.  The origin and driving mechanism
of these stellar jets are still major open issues, although the
presence of an accretion disk \citep{Livio2009pjc} and magnetic fields
are thought to be key in explaining the efficient jet acceleration and
collimation \citep{Blandford1982MNRAS,Cabrit2007LNP}.  There is a
jet-disk symbiosis, well established observationally in the case of
low-mass protostars, by which the surrounding accretion disk feeds the
jet by transporting gas and dust from the infalling envelope to the
protostar and the jet removes angular momentum and magnetic flux from
the disk allowing accretion to proceed \citep{Frank2014arXiv}.

High-mass stars ($M_\star>8\Msun$) form by accretion within massive
($\sim10^3\Msun$) and dense ($10^{4\text{-}5}$~cm$^{-3}$) molecular
clumps, with typical diameters of 1 pc and generally supported by
turbulent motions \citep{Garay2005IAUS,Zinnecker2007ARA&A,Tan2014arXiv}.  These
clumps harbor the luminous embedded infrared sources known as
high-mass young stellar objects (HMYSOs), that represent an early
evolutionary stage of a single high-mass star or a multiple stellar
system.  {It is probable, also, that some high-mass stars are born
  forming tight binary systems that will eventually merge, populating
  the highest end of the stellar mass spectrum \citep{Sana2012Sci}.}
Toward some {massive} clumps, the following phenomena ---
analogous to the ones observed in low-mass star formation --- are
detected:
\begin{itemize}
\item{Extended infalling envelopes, with inflows
  motions involving a sizable fraction of the molecular clump
  \citep{Ho1986ApJ,Zhang1997ApJ,Garay2002ApJ,Garay2003ApJ,Wu2003ApJ}. }
\item{Bipolar molecular outflows, which are poorly collimated but much more
  massive and energetic than in the low-mass case
  \citep{Zhang2001ApJ,Beuther2002AA,Wu2004AA,Zhang2005ApJ,Beltran2011AA}.}
\item{Rotation-flattened molecular structures surrounding the HMYSOs,
  ranging from transient toroids with sizes $\ge5000$
    AU \citep[][]{Zhang2005IAUS,Beltran2006Natur} to more stable, disklike
  structures of $500$-$2000$ AU where centrifugal support
  {may play a role}
  \citep{Patel2005Natur,Franco-Hernandez2009ApJ,Qiu2012ApJ,Sanchez-Monge2013AA,Beuther2013AA,Hunter2014ApJ}. }
\end{itemize}
{When} both bipolar outflows and rotating molecular structures are
detected, the symmetry axis of the former and the velocity gradient of
the latter are usually perpendicular.  Most disklike structures around
HMYSOs are, however, considerably different compared to low-mass
circumstellar disks: they are not thin or supported entirely by
rotation, and they might be unstable to further fragmentation.  The
closest analogs to low-mass circumstellar disks are the ones
discovered around HMYSOs that are not deeply embedded
\citep[e.g.,][]{Kraus2010Natur,Fallscheer2011ApJ}.  In addition, young
high-mass stars emit copious amounts of UV radiation that ionize their
surroundings \citep{Keto2006ApJ,Keto2008ApJ109}.  Important unsettled
questions are: Does the infall extend all the way to the molecular
core? Do accretion disks exist within rotating cores?  Are molecular
outflows driven by {underlying collimated jets} powered by
accretion?  {Numerical models have shown that disks
  \citep[e.g.,][]{Kuiper2011ApJ} and jets
  \citep[e.g.,][]{Vaidya2011ApJ} are theoretically possible to form
  and sustain around young high-mass stars.}

%% There are currently {a few} examples of high-mass young stellar objects
%% with bolometric luminosity $L_{\rm bol} > 50,000~\Lsun$ associated
%% with an infalling envelope, a collimated jet, and a bipolar molecular
%% outflow, {among them}: G343.1262$-$00.0620 \citep[also IRAS
%%   16547$-$4247]{Garay2003ApJ,Garay2007AA}, G345.4938+01.4677
%% \citep[also IRAS 16562$-$3959]{Guzman2010ApJ,Guzman2011ApJ} and {IRAS 13481$-$6124 
%% \citep{Kraus2010Natur}}.  
%% In  G343.1262$-$00.0620 and G345.4938+01.4677 (G345.49+1.47 hereafter), 
%% ionized lobes are detected aligned with the collimated wind.  These
%% lobes are analogous to Herbig-Haro objects, and reveal the presence of
%% fast ionization shocks.  {G343.1262$-$00.0620 
%% also has a disklike} molecular structure rotating around the jet axis
%% \citep{Franco-Hernandez2009ApJ}.

Optical and radio continuum observations indicate that a fraction of
the matter in the young stellar jet is in the form of ionized gas
\citep{Anglada1996ASPC}, i.e., the ionized jet.  {Examples of
  ionized jets associated to HMYSOs observed in radio continuum are
  IRAS 18162$-$2048 \citep[also HH 80-81,][]{Marti1993ApJ}, Cepheus A HW2
  \citep{Rodriguez1994ApJ}, IRAS 20126+4104 \citep{Tofani1995AAS,Cesaroni1997AA},
  G192.16$-$3.82 \citep{Shepherd1998ApJ}, W75N VLA 3
  \citep{Carrasco-Gonzalez2010AJ}, AFGL 2591 VLA 3
  \citep{Johnston2013AA}, G35.2$-$0.7N \citep{Gibb2003MNRAS}, NGC7538
  IRS 1 \citep{Sandell2009ApJ}, G343.1262$-$00.0620 \citep[also IRAS
    16547$-$4247]{Garay2003ApJ,Garay2007AA}, and G345.4938+01.4677
  \citep[also IRAS 16562$-$3959]{Guzman2010ApJ,Guzman2011ApJ}.} Most
of the present knowledge about ionized jets comes from studies at
optical and near-infrared (NIR) wavelengths of low-mass young stars
still associated with their protostellar disks, but no longer embedded
in their parental molecular cores, referred to as Class II objects
\citep{Ray2007PrPl}.  Young stars in earlier evolutionary phases are
still deeply embedded within their parental cores of dust and gas and
are thus undetectable at optical or NIR observations. This is the
general {situation in jets} associated with high-mass stars. Radio
continuum observations, on the other hand, are not affected by dust
absorption and are able to probe the characteristics of deeply
embedded ionized jets.

Physical parameters of ionized jets {associated with high-mass
  stars, such as the degree of collimation}, ionization fraction, or
kinematics, are not well determined, and they are usually constrained
from observations of the lobes.  In particular, estimates of the
velocity of the gas in the jet are derived from measurements of the
lobes proper motion.  In most cases, the estimates are close to
$\sim$500~\kms \citep{Marti1998ApJ,Curiel2006ApJ,Rodriguez2008AJ}, 
considerably faster than the jet velocity of their
low-mass counterparts. The dominant assumption in the literature has
been that the velocity of the ionized gas within the jet is similar to
that of the lobes.  Until now, the only direct observational support
for this assumption {has been} provided by hydrogen recombination
line (HRL) observations made by \citet{Jimenez-Serra2011ApJ} toward
the B-type YSO Cepheus A HW2.  While HRLs have become standard tools
to study regions of ionized gas
\citep{Mezger1968Sci,Brown1978ARA&A,Gordon2002ASSL}, this has not been
{true} for jets, which have much weaker flux densities and smaller
sizes than classical \hii\ regions.

We present ALMA  {Band-3} observations of
the HMYSO  {G345.4938+01.4677 (also IRAS 16562$-$3959). This HMYSO is
associated with an ionized wind and symmetrically located lobes
detected in centimeter radio continuum by \citet{Guzman2010ApJ},  
an infalling envelope, and 
a bipolar molecular outflow \citep{Guzman2010ApJ,Guzman2011ApJ}.}  
In the
following, we use the name IRAS 16562$-$3959 to refer to the more
extended, $\sim40\arcsec$ angular size molecular clump characterized
by single dish observations
\citep[e.g.,][]{Faundez2004AA}. G345.4938+01.4677 (G345.49+1.47 hereafter) 
is then the central
dominating HMYSO within  IRAS 16562$-$3959.  For the
present work, we adopt a distance to IRAS 16562$-$3959 of 1.7 kpc
\citep{Lopez2011AA}.\footnote{Note that \citet{Lumsden2013ApJS}
  derived a spectro-photometric distance of 2.4 kpc.} 
{
The gas mass of IRAS 16562$-$3959 is $\sim 900\Msun$. Assuming that
approximately 30\% of this mass will end up as stellar mass 
in a cluster \citep{Lada2003ARA&A}, and using the empirical 
relationship 
$M_{\rm max}=1.2 M_{\rm cluster}^{0.45}$ where $M_{\rm max}$ is the mass of the 
most massive member of a cluster and $M_{\rm cluster}$ is the cluster's mass 
\citep[in units of solar mass, see][]{Larson2003ASPC}, we determine that the likely 
mass of the central star of  G345.49+1.47 is $\sim15\Msun$.
In this work, we focus on results derived from the continuum,
  HRLs, and sulfuretted molecular lines, and we leave the analysis of
  other observed molecular tracers (e.g., SiO, CH$_3$OH, and C$_2$H) for
  upcoming publications.}

Section 2 presents the ALMA observations toward G345.49+1.47 and data
reduction.  Section 3 presents the results from the radio continuum, 
from the three HRLs, and from the sulfuretted molecules. We
discuss and model the results in Section 4, where we suggest that
the G345.49+1.47 radio continuum and HRL emission are best explained as
arising from a photo-ionized disk wind. Section 5 summarizes {our} main
conclusions.

\section{OBSERVATIONS}

\begin{deluxetable}{lcccc}
%\rotate
   \tablewidth{0pc} \tablecolumns{5} \tabletypesize{\small}
   \tablecaption{Spectral Setting and Angular  Resolution\label{tab-SpWs}}
   \tablehead{ \colhead{~} & \colhead{85.4~GHz}&\colhead{87.2~GHz}&\colhead{97.6~GHz}&\colhead{99.3~GHz}}
\startdata
Spectral Window Limits (GHz) & [84.42, 86.30] & [86.18, 88.06] &[96.68, 98.56] & [98.38, 100.26] \\
Synthesized Beams  & 2.51\arcsec$\times$1.42\arcsec& % 
2.47\arcsec$\times$1.40\arcsec& 2.22\arcsec$\times$1.24\arcsec &  %
2.18\arcsec$\times$1.26\arcsec \\
Position Angle & $97.8\arcdeg$ & $97.3\arcdeg$ & $97.2\arcdeg$ & $97.7\arcdeg$ \\
\enddata
\end{deluxetable}

Data were obtained using the {\it Atacama Large Millimeter/sub-millimeter Array} (ALMA) 
during Cycle 0 using the extended array
configuration {(longest and shortest baselines 453 and 21 m, respectively)}. We observed G345.49+1.47
{for} $\sim$188 minutes {on-source} in Band-3, which covers the 3 mm atmospheric
window, in 5 scheduling blocks. Two scheduling blocks were observed with 17 12-m
antennas, and the other three with 25.  The phase center of the {array was}
 R.A.$=16^{\rm h}59^{\rm m}$41\fs63,
decl.$=-$40\arcdeg03\arcmin43\farcs61 (J2000), the
position of the central jet source identified by \citet{Guzman2010ApJ}.

The observations covered {four} spectral windows (SpWs), each one spanning
1.875 GHz.  
Each SpW {consisted} of 3840 channels of 488~kHz width and
were centered at 85.4, 87.2, 97.6, and 99.3~GHz. We  use these
frequencies to refer to each SpW throughout this work.  The effective
spectral resolution is approximately two times the channel width, that
is 976~kHz, which corresponds to $\sim$3.0~\kms.  Table \ref{tab-SpWs}
gives the spectral limits of each SpW, and the synthesized beam of the
array at these frequencies, which {was} typically $2\farcs3\times1\farcs3$ with
a position angle of 97\arcdeg\ (Table \ref{tab-SpWs}).  The primary
beam FWHM {was} $67$\arcsec, and the typical system temperature was $70$~K.

We recalibrated the data using the Common Astronomy Software
Applications (CASA)  \citep[v.4.0.1][]{Petry2012ASPC}.  The sources
Neptune and Titan were used as flux
calibrators,\footnote{Butler-JPL-Horizons 2012 flux model.} J1924$-$292
and 3C279 were used as bandpass calibrators, and J1717$-$337 was used
as a gain calibrator.  The {flux densities} derived for J1717$-$337 were
$1.46\pm0.03$, $1.44\pm0.03$, $1.37\pm0.03$, and $1.36\pm0.03$~Jy at
85.4, 87.2, 97.6, and 99.3~GHz, respectively.

Maps were generated by Fourier transformation of the robust-weighted
visibilities \citep{Briggs1995PhD}, robust parameter $=0$, and
deconvolved using the {\tt clean} task within CASA.   
The pixel size used was
$0\farcs3\times0\farcs3$\ in all cases.  
Most of the flux {density} detected
toward G345.49+1.47 arises from  a compact continuum source of
$\sim0.1$~Jy, which allowed us to perform an additional
phase self-calibration iteration.  All visibilities were calibrated in
phase using this self-calibrated solution.

The spectral location of HRLs and strong molecular lines {was} masked
out by visual inspection to {isolate} the continuum {emission}.  {This continuum} was
subtracted from the visibility data using the CASA task {\tt
  uvcontsub}.  The noise level achieved in the continuum maps for each
{SpW was} typically 50~$\mu$Jy~beam$^{-1}$, as measured by the rms
of the final images.  Deconvolved images per channel were obtained for
selected spectral lines using \texttt{clean}, achieving noise levels
of 1-2~mJy~beam$^{-1}$. The continuum and spectral cube noise levels attained
are comparable to the theoretical sensitivity calculated using the
ALMA Observing Tool, {which} indicates 25~$\mu$Jy per SpW for the
continuum and 1.4~mJy per channel for the spectral lines.

Fully reduced datasets, continuum, and spectral cubes are 
 publicly available through the 
{Dataverse}.\footnote{http://dx.doi.org/10.7910/DVN/24060}

{\section{RESULTS}\label{sec-Results}}

{\subsection{Continuum emission}\label{sec-Cont}} 

\begin{figure}
\includegraphics[width=\textwidth]{{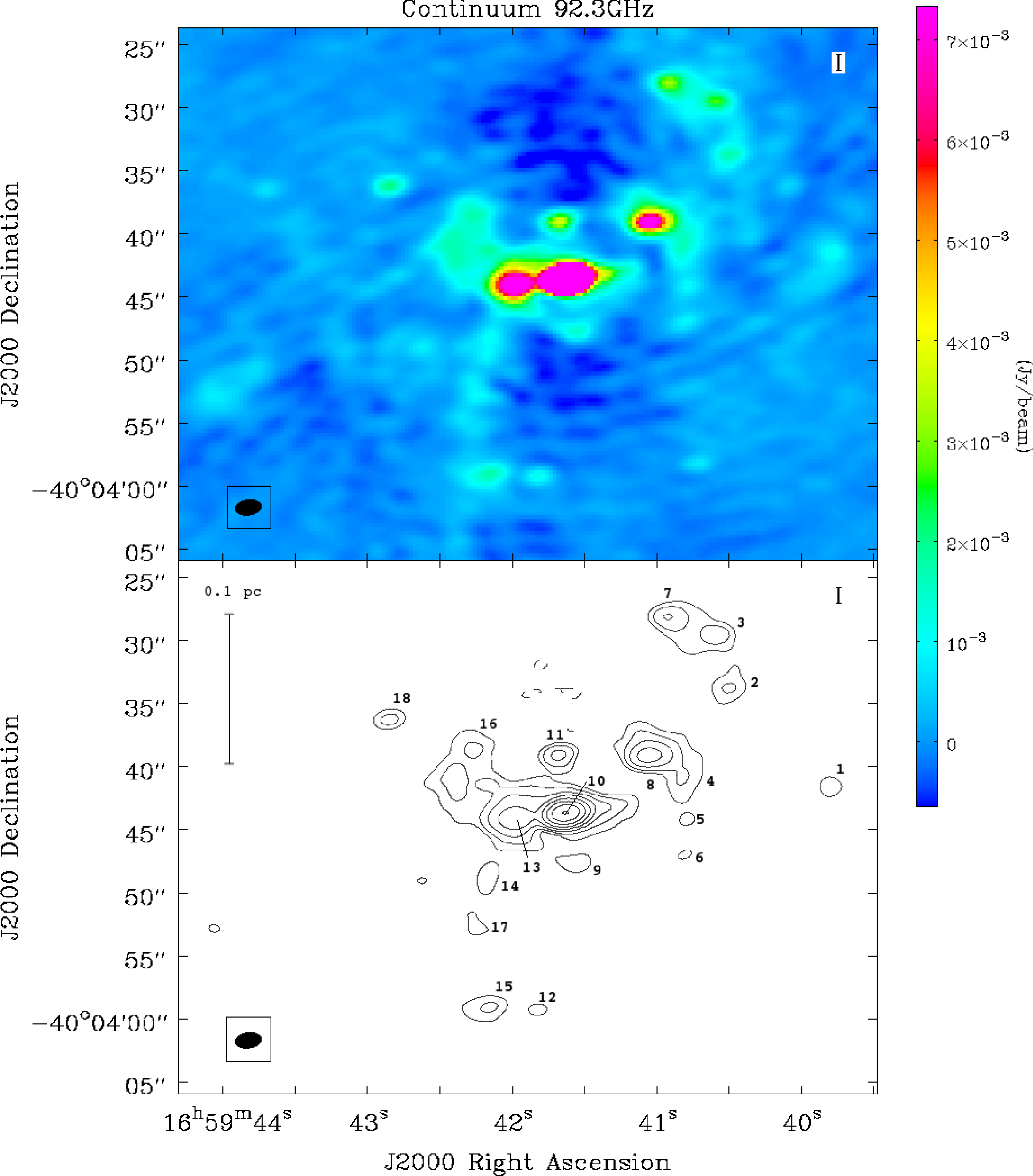}}%
\figcaption{Continuum emission detected at 92.3~GHz toward IRAS
  16562$-$3959. {\it Bottom panel:} Contours shown correspond to $-15,
  15, 30, 60, 120, 250, 500, 1000$, and 2000\,$\sigma$, with
  $\sigma=50~\mu$Jy~beam$^{-1}$.  {$\sigma$ represents the
    r.m.s.\ variations of the dynamic range limited image measured at
    the edge of the field.}  Compact sources are identified by the
  number given in Table \ref{tab-ClumpSou}.  All sources are within
  the primary beam. \label{fig-contIm}} 
\end{figure}

Figure \ref{fig-contIm} shows the {92 GHz} deconvolved map of the continuum
emission obtained combining the four SpWs (Table \ref{tab-SpWs}). The
four SpW maps display similar morphology.  Figure
\ref{fig-contIm} shows that the emission is dominated by a central,
bright compact source (Source 10) associated with G345.49+1.47, with an
integrated flux density of $\sim0.1$~Jy. We also distinguish 17
additional compact secondary sources, and  extended emission
partially recovered by the interferometer.

\begin{deluxetable}{lccrrrrrr}
\rotate
   \tablewidth{0pc} \tablecolumns{9} \tabletypesize{\small}
   \tablecaption{Continuum Sources Detected toward IRAS 16562$-$3959\label{tab-ClumpSou}}
   \tablehead{% 
     \colhead{Source} & \colhead{R.A.}&\colhead{Decl.} & \multicolumn{4}{c}{Flux Density}&\colhead{Spectral}&\colhead{$\chi^2_r$}\\\cline{4-7}%
     \colhead{~} &\colhead{(J2000)}&\colhead{(J2000)} & \colhead{85.4~GHz}&\colhead{87.2~GHz}&\colhead{97.6~GHz}&\colhead{99.3~GHz}&\colhead{Index\tablenotemark{a}}&\colhead{~}\\%
     \colhead{~}&\colhead{}&\colhead{}&\colhead{(mJy)}&\colhead{(mJy)}&\colhead{(mJy)}&\colhead{(mJy)}&\colhead{~}&\colhead{~}}
\startdata
1 & 16$^{\rm h}$59$^{\rm m}$39\fs81 &$-$40\arcdeg03\arcmin41\farcs4 & 2.53(0.13) & 2.55(0.13) & 2.71(0.14) & 2.61(0.14) & 0.33(0.6) & 0.2 \\
2 & 16 59 40.52 &$-$40 03 33.7 & 2.64(0.11) & 3.58(0.12) & 4.55(0.12) & 4.7(0.12) & 3.08(0.4) & 11.0 \\
3 & 16 59 40.60 &$-$40 03 29.5 & 4.67(0.12) & 5.3(0.12) & 7.71(0.13) & 8.15(0.13) & 3.54(0.2) & 1.3 \\
4 & 16 59 40.83 &$-$40 03 40.8 & 4.72(0.10) & 5.16(0.11) & 7.57(0.11) & 7.9(0.11) & 3.4(0.2) & 0.4 \\
5 & 16 59 40.81 &$-$40 03 44.2 & 1.89(0.10) & 2.1(0.11) & 2.55(0.11) & 2.56(0.11) & 1.89(0.5) & 0.6 \\
6 & 16 59 40.80 &$-$40 03 46.9 & 1.69(0.11) & 1.8(0.11) & 2.04(0.11) & 1.94(0.11) & 0.98(0.7) & 0.5 \\
7 & 16 59 40.90 &$-$40 03 28.0 & 5.55(0.12) & 6.18(0.12) & 9.07(0.12) & 9.14(0.13) & 3.31(0.2) & 4.2 \\
8 & 16 59 41.06 &$-$40 03 39.1 & 13.54(0.10) & 14.7(0.10) & 21.31(0.1) & 22.39(0.1) & 3.31(0.07) & 1.4 \\
9 & 16 59 41.55 &$-$40 03 47.8 & 1.62(0.10) & 1.82(0.10) & 2.13(0.1) & 2.23(0.1) & 1.85(0.6) & 0.5 \\
10\tablenotemark{b} & 16 59 41.63 & $-$40 03 43.6 & 103.8(0.90) & 105.7(1.0) & 118.5(1.1) & 120.8(1.2) & 1.01(0.1) & 0.0 \\
11 & 16 59 41.68 &$-$40 03 39.1 & 3.34(0.10) & 3.74(0.10) & 4.87(0.1) & 5.18(0.1) & 2.69(0.3) & 1.2 \\
12 & 16 59 41.82 &$-$40 03 59.3 & 0.69(0.11) & 0.69(0.11) & 1.31(0.12) & 1.32(0.12) & 4.8(1.5) & 0.3 \\
13\tablenotemark{b} & 16 59 41.99 &$-$40 03 43.9 & 26.24(2.02) & 27.26(2.0) & 37.44(2.02) & 37.46(2.02) & 2.52(0.8) & 0.2 \\
14 & 16 59 42.16 &$-$40 03 48.3 & 2.17(0.10) & 2.27(0.10) & 3.33(0.1) & 3.44(0.1) & 3.18(0.5) & 0.2 \\
15 & 16 59 42.17 &$-$40 03 59.1 & 3.13(0.12) & 3.26(0.12) & 5.21(0.12) & 5.61(0.12) & 3.99(0.3) & 0.3 \\
16& 16 59 42.29 &$-$40 03 38.3 & 3.24(0.10) & 3.56(0.10) & 5.3(0.11) & 5.32(0.11) & 3.33(0.3) & 2.1 \\
17 & 16 59 42.27 &$-$40 03 52.9 & 1.62(0.11) & 1.94(0.11) & 3.08(0.11) & 3.1(0.11) & 4.14(0.6) & 1.6 \\
18& 16 59 42.85 &$-$40 03 36.2 & 2.17(0.11) & 2.18(0.11) & 2.74(0.12) & 2.51(0.12) & 1.35(0.4) & 1.5 \\
\enddata
\tablecomments{Fluxes are corrected for primary beam response.}
\tablenotetext{a}{See \S 3.1.}
\tablenotetext{b}{Peak position  of the 2-D Gaussian fit.}
\end{deluxetable}

The continuum sources identified in Figure \ref{fig-contIm} correspond
to compact emission detected above 0.75~mJy in the combined continuum
map.  {This threshold is $\ge15~\sigma$, where $\sigma$ is the
r.m.s.\ variations of the image measured at the edge of the field.  This number
does not represent random noise, but it arises mainly because of dynamic
range limitations of the image.}
All sources, except perhaps source 18,
are embedded in somewhat extended and diffuse emission.

The positions of the continuum sources are given in Table
\ref{tab-ClumpSou}.  The coordinates  correspond to the 
position of the emission peak
determined by using the CASA function {\tt maxfit}, except for Sources 10
and 13, for which we fit two 2-D Gaussians using the CASA task {\tt
  imfit}.  For each source, the coordinates determined in the four
SpWs are consistent within 0\farcs15.  Note that the position of
Source 10 corresponds to the position of the jet source reported by
\citet{Guzman2010ApJ}.  The deconvolved size of Source 10
is $\lesssim$0\farcs4 in each SpW, which is less than one-third of the
beam size. {Since} Source 10 is not completely isolated,
we refrain from further analyzing its deconvolved size here.

Table \ref{tab-ClumpSou} also lists the integrated flux densities in
each of the SpWs and the derived in-band spectral index.  For Sources
10 and 13, integrated flux densities are derived from the Gaussian
fittings. For the rest of the sources, we integrated the flux density
within boxes of two times the size of the beam.  The uncertainty
assumed for each flux density is either $2\sigma\sim0.1$~mJy or the
uncertainty derived from the 2-D Gaussian fit.  Best-fit spectral
indexes and their uncertainties were obtained by weighted
least-squares (or $\chi^2$) minimization using the procedure described
in \citet{Lampton1976ApJ}. Unless stated otherwise, this is the
procedure we follow throughout this work. \chg{The last column of
  Table \ref{tab-ClumpSou} shows $\chi^2_r$, that is, the least-square
  value divided by the number of degrees of freedom (2).}

{\subsection{HRL emission}\label{sec-HRL}}

HRL emission was detected toward 
the central Source 10  in three transitions: 
\ha{40}, \ha{42}, and \hb{50}, whose rest frequencies
are 99022.95, 85688.39, and 86846.96 MHz, respectively. 
The spatial distribution of the emission of the three HRLs
is similar in all velocity channels and can be described as an
unresolved source located within 0\farcs1 with respect to the phase center.

\begin{figure}
\centering\includegraphics[width=1.1\textwidth]{{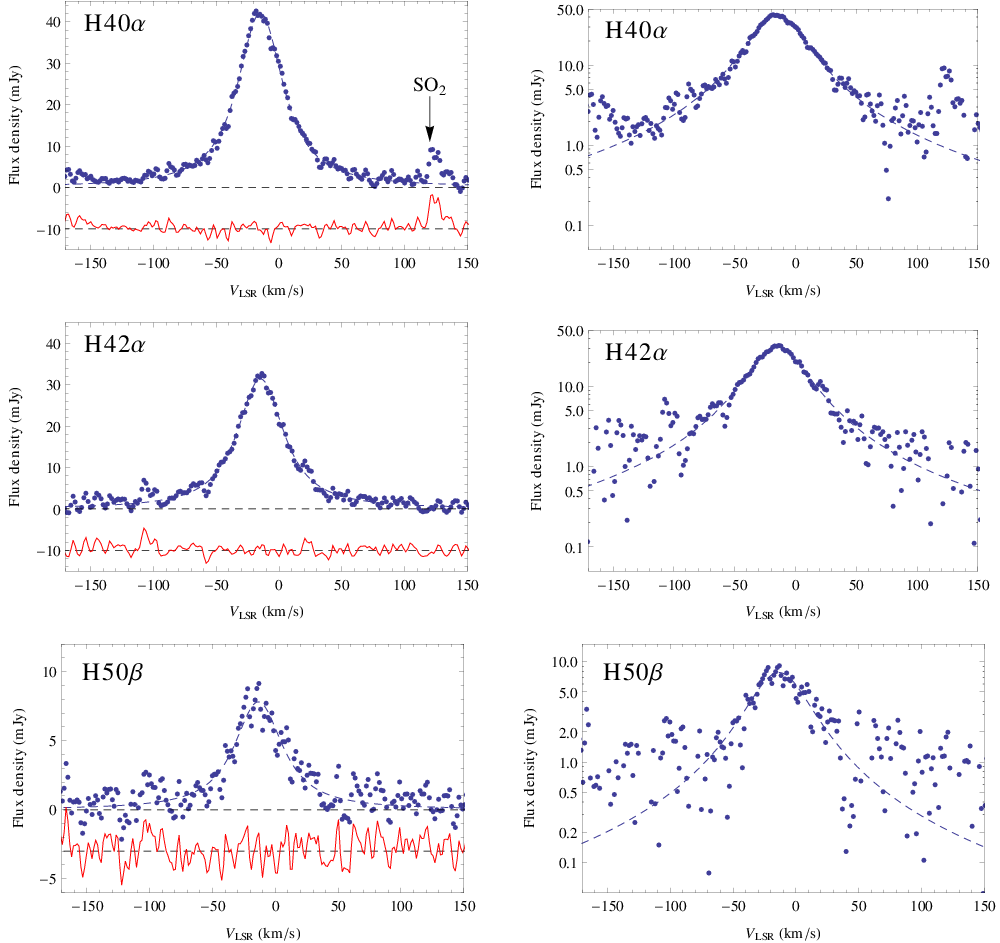}}
\figcaption{Spectra of the HRLs detected toward
  G345.49+1.47.  Blue dots: Observed data. Dashed lines: Voigt profile
  fits to the spectra, using the same central velocity and thermal
  parameter for the three lines.  \emph{Left column:} {The flux}
  density plotted in linear scale. The red lines indicate the
  residuals from the fits.  \emph{Right column:} {The flux density}
  displayed in log scale in order to emphasize the wing
  emission.\label{fig-HrlsVoigt}}
\end{figure}

Figure \ref{fig-HrlsVoigt} shows the spectra of the three HRLs
detected toward G345.49+1.47.  Table \ref{tab-Desc} gives the observed
parameters of the line profiles at the peak position.  From this table
and Figure \ref{fig-HrlsVoigt} it is evident that the HRL profiles
exhibit extended wing emission.  Two further characteristics in the
observed spectrum are worth mentioning: {First,  the 
line located at a velocity of $\sim+125$~km~s$^{-1}$ in the \ha{40}
spectra. This line  is narrower than the HRLs,  and was identified as  one of the  
SO$_2$ rotational transitions (see \S \ref{sec-Slines}).}
{Second,} an unidentified feature appears in the spectrum of the
\ha{40} toward $V_{\rm LSR}<-140$~\kms. We have excluded the velocity
range affected by this emission in the analysis of the HRLs.

\begin{deluxetable}{lccccc}
%\rotate
   \tablewidth{0pc} \tablecolumns{6} \tabletypesize{\small}
   \tablecaption{Observed parameters of the HRLs\label{tab-Desc}}
   \tablehead{ 
     \colhead{}&\colhead{Peak Flux}&\colhead{$V_{\rm LSR}$ of}&\colhead{FWHM}&\colhead{FWZP\tablenotemark{a}} &  \colhead{Integrated}\\%
     \colhead{}&\colhead{Density}    &\colhead{Peak} &\colhead{(\kms)}&\colhead{(\kms)}& \colhead{flux}\\%
     \colhead{}&\colhead{(mJy)}         &\colhead{(\kms)} &\colhead{}&\colhead{}&\colhead{(Jy~\kms)} } 
\startdata
H40$\alpha$  &  42.6(1) & $-$18.3 & 44.4 & 360   &2.86(0.05)  \\
H42$\alpha$   &  32.7(1) & $-$13.6 & 39.3 & 227  &1.95(0.04)  \\  
H50$\beta$    &  9.1(1) & $-$13.2 & 50.2 & 124   &0.43(0.03) \\  
\enddata
\tablenotetext{a}{Full width at zero power.}
\end{deluxetable}

{\subsection{Sulfuretted molecules}\label{sec-Slines}}

Table \ref{tab-Slines} shows the list of sulfuretted molecules
detected in our {observations toward  G345.49+1.47 and  summarizes their main characteristics.}  
{The detected species}  are sulfur monoxide (SO) and its
$^{34}$S isotopologue, sulfur dioxide (SO$_2$), carbonyl sulfide
(OCS), and carbon monosulfide (CS) and its $^{33}$S
isotopologue. 
We synthesized spatial maps of the emission {for} all the lines, except 
of SO$_2$ $28_{7,21}\rightarrow29_{6,24}$ (see next section).
Columns 2-4 {of Table \ref{tab-Slines}} list the frequency, transition, and
energy associated with the upper level of the transition in K
($E_u=kT$) obtained from  
the JPL \citep{Pickett1998JQSRT} 
and CDMS \citep{Muller2001AA} databases 
consulted through the {Splatalogue}\footnote{http://www.cv.nrao.edu/php/splat}  \citep{Remijan2007AAS}.  
Column 5 indicates whether  we detect a
velocity gradient associated with the central compact component.  
Column 6 gives  the velocity
integrated from $V_{\rm LSR}$ $-30$ to 0 ~\kms\ line flux 
within a   5\arcsec$\times$5\arcsec\ box 
centered on  G345.49+1.47 
(equatorial orientation).  
Finally, columns 7-9 display the results of Gaussian
fittings to the line flux integrated in the same central
5\arcsec$\times$5\arcsec\ region.  The Gaussian parameters roughly
describe the most important characteristics of each line, but we
stress that Gaussian models to the line profiles are generally poor.
Additionally, since the spectral resolution of our data is
$\sim$3.0~\kms, in the majority of the lines, we {have only two} 
independent spectral sampling points per FWHM (column 9 of
Table \ref{tab-Slines}). This sampling is too scarce to attempt a
detailed modeling of the line profiles.

\begin{deluxetable}{lrlrccccc}
\rotate
   \tablewidth{0pc} \tablecolumns{9} \tabletypesize{\small}
   \tablecaption{Sulfuretted molecular species detected.\label{tab-Slines}}
   \tablehead{ \multicolumn{6}{c}{~} & \multicolumn{3}{c}{Gaussian Fitting}\\
     \cline{7-9}
     \colhead{Species} & \colhead{Rest Frequency} & \colhead{Transition} &\colhead{$E_u/k$} &  \colhead{Velocity} & \colhead{Line Flux ($W$)} &  \colhead{Peak}& \colhead{$V_{\rm LSR}$} & \colhead{FWHM} \\% 
     \colhead{~}&\colhead{(MHz)}&\colhead{~}&\colhead{(K)}&\colhead{Gradient}&\colhead{(Jy~\kms)}& \colhead{(Jy)}& \colhead{(\kms)} & \colhead{(\kms)}\\%
     \colhead{(1)}&\colhead{(2)}&\colhead{(3)}&\colhead{(4)}&\colhead{(5)}&\colhead{(6)}&  \colhead{(7)}& \colhead{(8)} & \colhead{(9)} \\}
   \startdata
SO        & 86093.950   & $J_N=2_2\rightarrow1_1$     & 19.3              & Yes & 6.22            & 0.873 & $-$12.9  & 6.6 \\	  
          & 99299.870   & $J_N=3_2\rightarrow2_1$      & 9.2              &  Yes & 14.1            & 1.777 & $-$13.8 & 7.2 \\		     
          & 100029.64\phn& $J_N=4_5\rightarrow4_4$     & 38.6             &  Yes & 2.39            & 0.544 & $-$13.2 & 6.5 \\	 
$^{34}$SO & 96781.76\phn & $J_N=4_5\rightarrow4_4$   & 38.1               &  Yes &0.099             & 0.016 & $-$13.7 & 5.4 \\	 
          & 97715.317   & $J_N=3_2\rightarrow2_1$      & 9.1              &  Yes & 1.30            & 0.196 & $-$12.9 & 6.0 \\	 
SO$_2$    & 86639.088   & $J_{K_pK_o}=8_{3,5}\rightarrow9_{2,8}$ & 55.2     &  Yes &0.97 & 0.118 & $-$13.1 & 7.5 \\	 
          & 86828.938   & $J_{K_pK_o}=20_{2,18}\rightarrow21_{1,21}$ &207.8 &  No& 0.025 & 0.002 & $-$10.6 & 9.4 \\	
          & 97702.334   & $J_{K_pK_o}=7_{3,5}\rightarrow8_{2,6}$ & 47.8     &  Yes& 1.32 & 0.170 & $-$12.9 & 7.3 \\	 
          & 98976.294\tablenotemark{a} 
                        & $J_{K_p,K_o}=28_{7,21}\rightarrow29_{6,24}$ &493.7 &  No &0.11 & 0.007 & $-$16.8 & 13.5 \\	
          & 99392.513   & $J_{K_pK_o}=29_{4,26}\rightarrow28_{5,23}$  & 440.7&  No &0.23 & 0.023 & $-$16.8 & 9.0 \\	 
OCS       & 85139.104 & $J=7\rightarrow6$    & 16.3                       &  No& 1.08& 0.130 & $-$14.1 & 6.0 \\	 
          & 97301.208 & $J=8\rightarrow7$  & 21.0     &  No& 1.63& 0.251 & $-$14.1 & 5.9 \\      
CS        & 97980.953 & $J=2\rightarrow1$ & 7.1      & No& 6.81 & 1.38 &$-$14.2  & 4.6 \\
C$^{33}$S  & 97172.09\phn & $J=2\rightarrow1$ & 6.3   & No& 0.19 & 0.052 &$-$13.5 & 3.5  \\
\enddata
\tablenotetext{a}{Data for this line was extracted from the \ha{42} spectrum after subtracting the Voigt model (see \S\,\ref{sec-HRL}).}
\end{deluxetable}

\begin{figure}
\includegraphics[width=\textwidth]{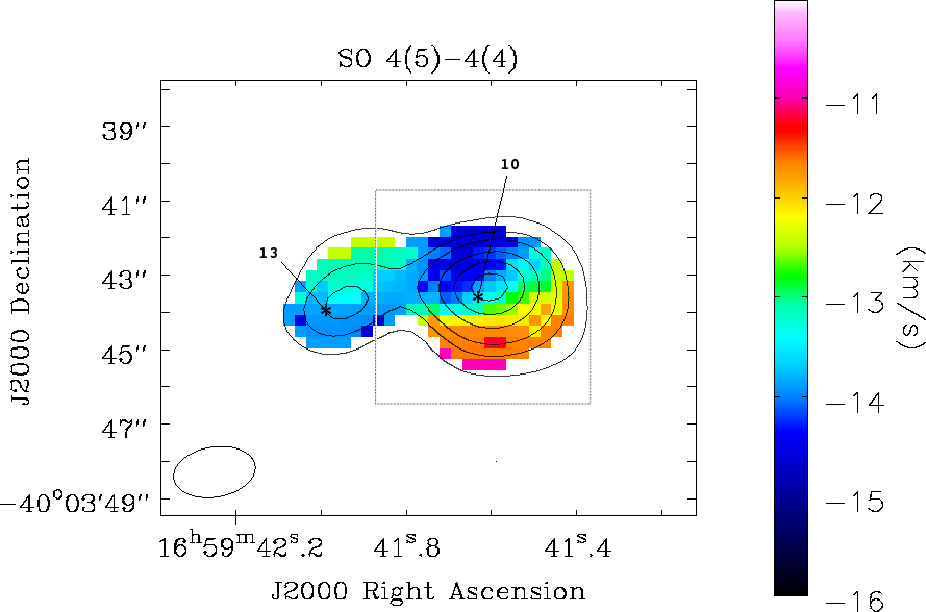}%
\figcaption{Moment 1 of the SO $J_N=4_5\rightarrow4_4$ emission
  detected toward G345.49+1.47 in color scale. Data below 20\% of the
  peak $=0.158$~Jy~beam$^{-1}$ {are} masked.  Black contours display the
  emission integrated in the velocity interval between $-22.5$ and
  $-2.5$~\kms. Contours correspond to 10, 20, 30, 50, 70, and 90\% of
  the peak value~$=1.16$ Jy beam$^{-1}$~\kms. The position of the
  continuum Sources 10 and 13 are indicated by  asterisks. The dashed square 
  indicates the 5\arcsec$\times$5\arcsec integration region used to calculate the line fluxes reported in
  column 6 of Table \ref{tab-Slines}.\label{fig-SO100m0m1}}
\end{figure}

The main morphological features of the integrated line maps (or
0th moment) associated with SO, $^{34}$SO, and SO$_2$ (the sulfur oxides)
are all similar. Taking SO
$J_N=4_5\rightarrow4_4$ as a representative example,
Figure~\ref{fig-SO100m0m1} shows contours of the velocity integrated flux
within [$-$30.0,0.0]~\kms.  Most of the emission comes from a 
central bright component  with a peak position
displaced about 0\farcs4 {northwest of Source 10}.
There is also emission from a weaker  source located $\sim4$\arcsec\ east of Source 10, consistent
with the position of Source 13.  From Figure \ref{fig-SO100m0m1} we also
see that the  selected  5\arcsec$\times$5\arcsec\ region  encloses well the
emission associated with G345.49+1.47, whose parameters for each line
are in columns 7-9 of Table \ref{tab-Slines}.

\begin{figure}
\includegraphics[width=\textwidth]{{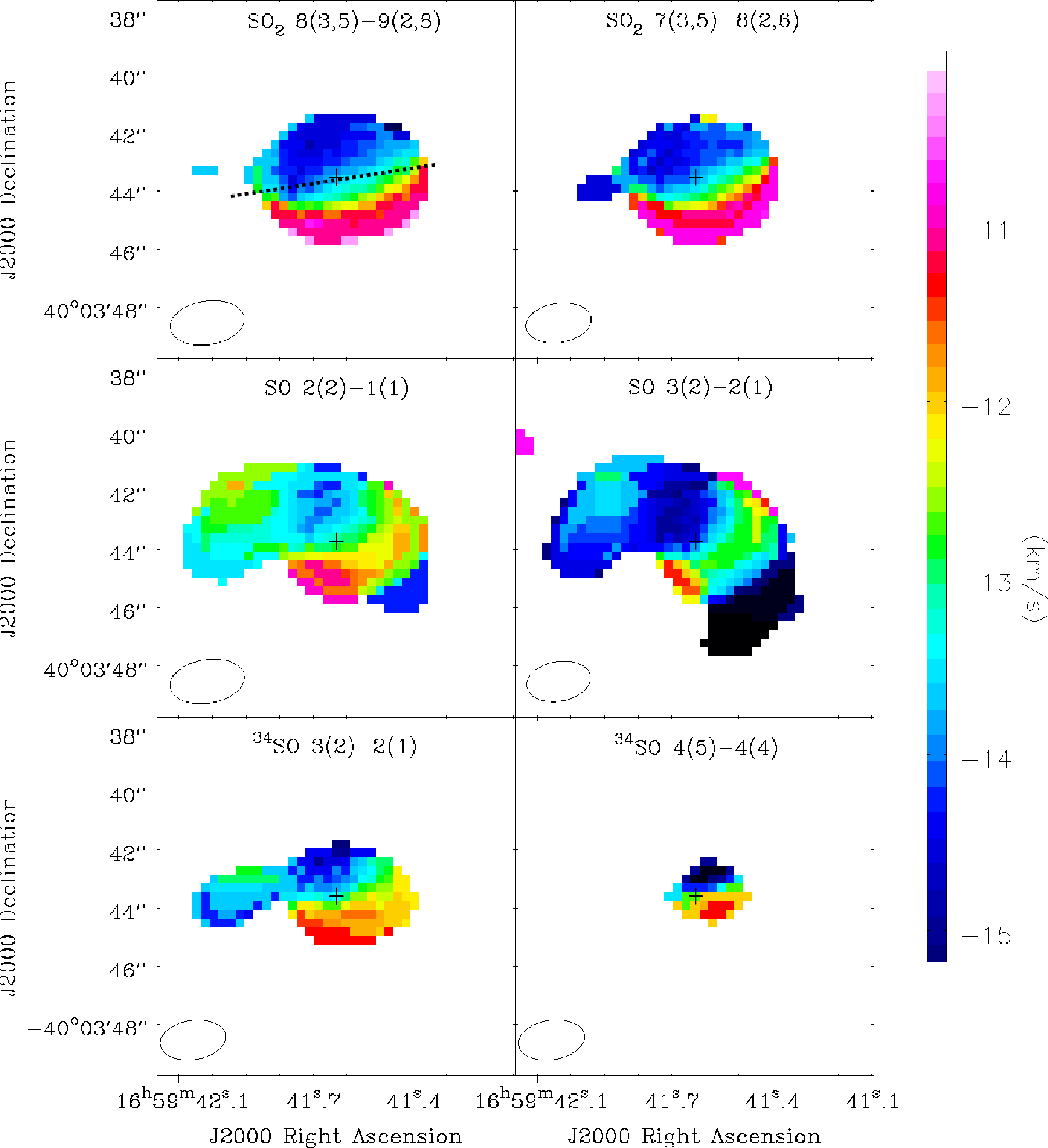}}%
\figcaption{{Moment-1}  maps of the sulfuretted molecules that
  display clear velocity gradients associated with G345.49+1.47. The
  color stretch is equivalent to that of Figure
  \ref{fig-SO100m0m1}. In each panel, a ``+'' sign indicates the
  continuum peak. {\it Top left panel:} The dashed black line
  corresponds to the direction inferred for the jet as traced by the
  ionized lobes \citep{Guzman2010ApJ}. \label{fig-OxoSrot}}
\end{figure}

Figures \ref{fig-SO100m0m1} and \ref{fig-OxoSrot} show the first
moment of {some of the} sulfur oxide lines.
There is a clear  velocity gradient
associated with G345.49+1.47.  The gradient directions and
magnitudes are similar in  all {these transitions}.

\begin{figure}
\includegraphics[width=\textwidth]{{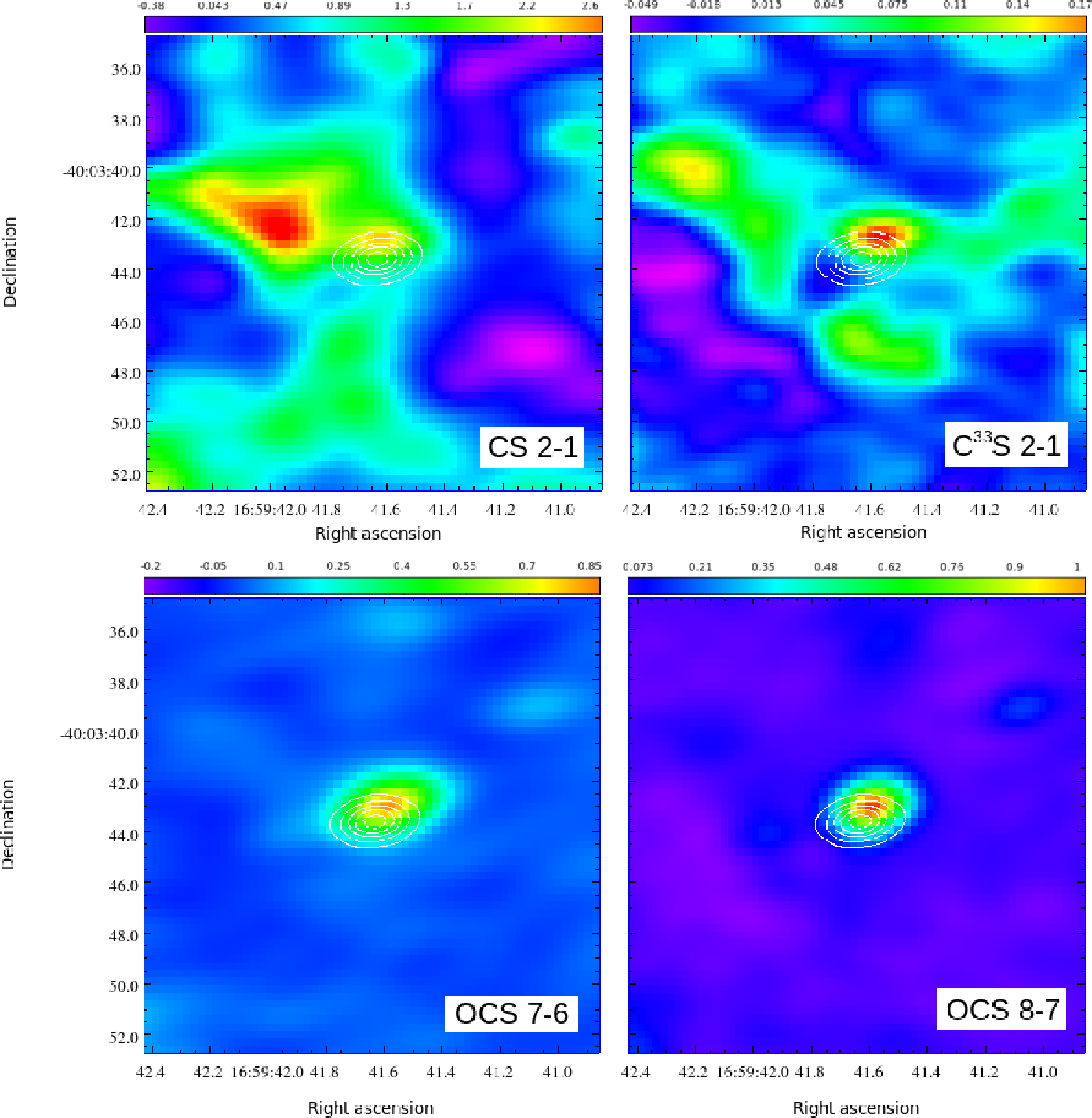}}%
\figcaption{Velocity integrated emission of the CS and OCS lines. The
  color bars indicate the scale in Jy~\kms.  White contours correspond to 
  continuum emission, dominated by Source 10, at 20, 40, 60, 80, and
  90\% of the peak (${\rm peak}=0.106$~Jy~beam$^{-1}$). \label{fig-OCS-CS-m0}}
\end{figure}

The morphology of the emission in the other sulfuretted species (CS, C$^{33}$S, and
OCS) is  different than that of the sulfur oxides.  Figure
\ref{fig-OCS-CS-m0} shows integrated line maps of the CS
2$\rightarrow$1, C$^{33}$S 2$\rightarrow$1, OCS 7$\rightarrow$6, and
OCS 8$\rightarrow$7 transitions.  The OCS emission is dominated by a
single compact source, but displaced from the center of the map
by $\sim$0\farcs7 to the northwest.  The CS and C$^{33}$S trace significant
extended emission in addition to a compact source near 
the map center but displaced $\sim1\farcs2$ to the northwest.  
This source is relatively more prominent in the C$^{33}$S
map {with} respect to the diffuse extended emission, when compared to CS.

\begin{figure}
\includegraphics[width=0.9\textwidth]{{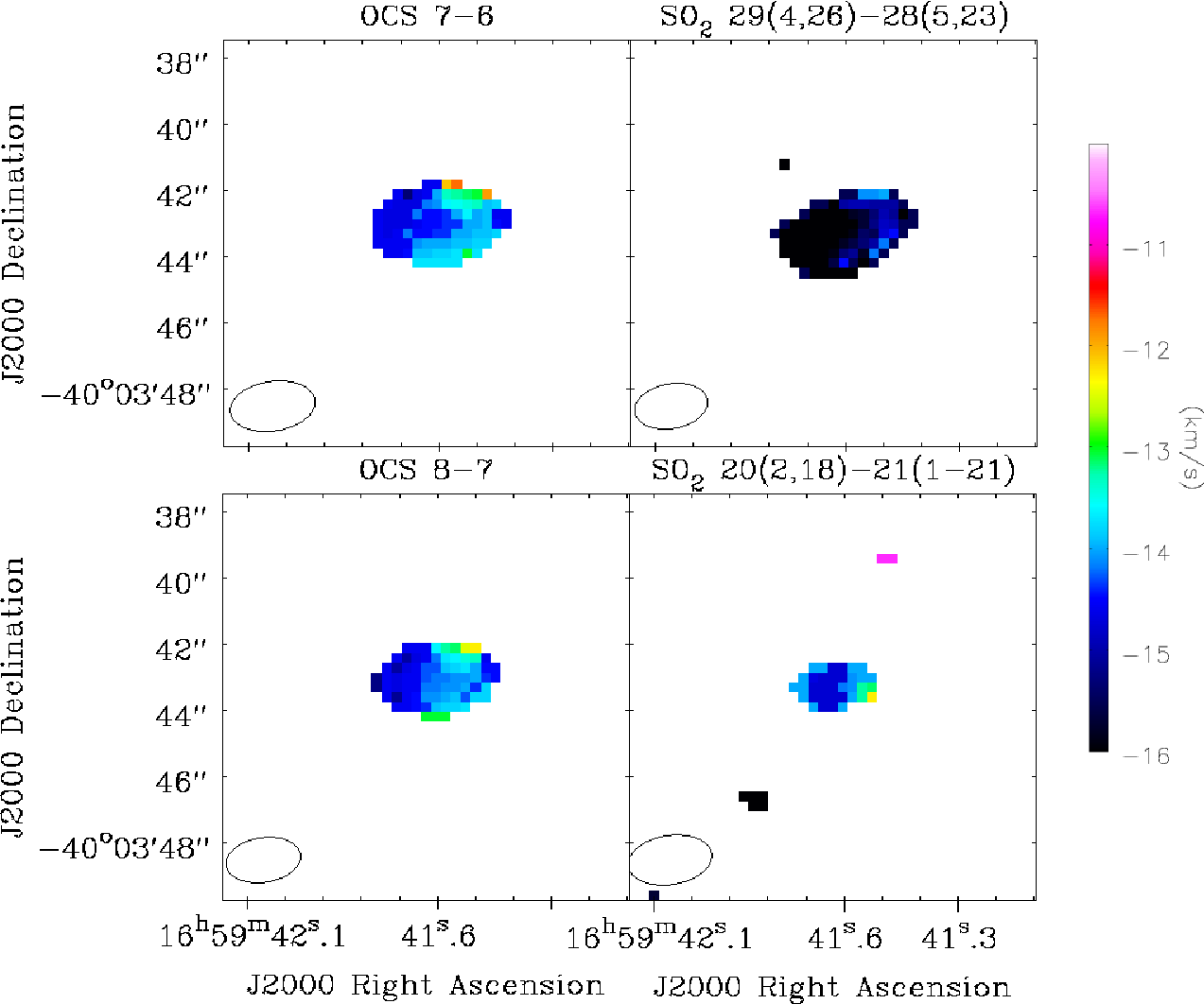}}%
\figcaption{Moment 1 of the sulfuretted molecular line emission
  associated with G345.49+1.47 that shows no velocity gradient
  ({except for} SO$_2$  $28_{7,21}\rightarrow29_{6,24}$). \label{fig-m1SnoRot}}
\end{figure}

{Figure \ref{fig-m1SnoRot} shows} first moment maps of the
OCS lines and of the SO$_2$ $20_{2,18}\rightarrow21_{1,21}$ and
$29_{4,26}\rightarrow28_{5,23}$ transitions.  
We do not detect any
velocity gradient in these lines.
Neither CS nor C$^{33}$S lines display velocity gradients analogous 
to the ones traced by the sulfur oxides.
{Finally, we identify emission from the  SO$_2$ $28_{7,21}\rightarrow29_{6,24}$ line 
near the \ha{40} HRL.  
We expect this SO$_2$ line to have an apparent velocity displacement of $+126$~km~s$^{-1}$ with
respect to the \ha{40} line, consistent with the observations 
(see \chg{previous} section and Figure \ref{fig-HrlsVoigt}). Having observed other strong SO$_2$
transitions strengthens this
identification.}

\section{DISCUSSION}
In this section, we discuss and analyze the continuum, {HRL}, and sulfuretted molecular line emission detected
toward the compact source G345.49+1.47 and the clump IRAS 16562$-$3959.

\subsection{Continuum sources toward IRAS 16562$-$3959}

\begin{figure}
\centering\includegraphics[height=0.8\textheight]{{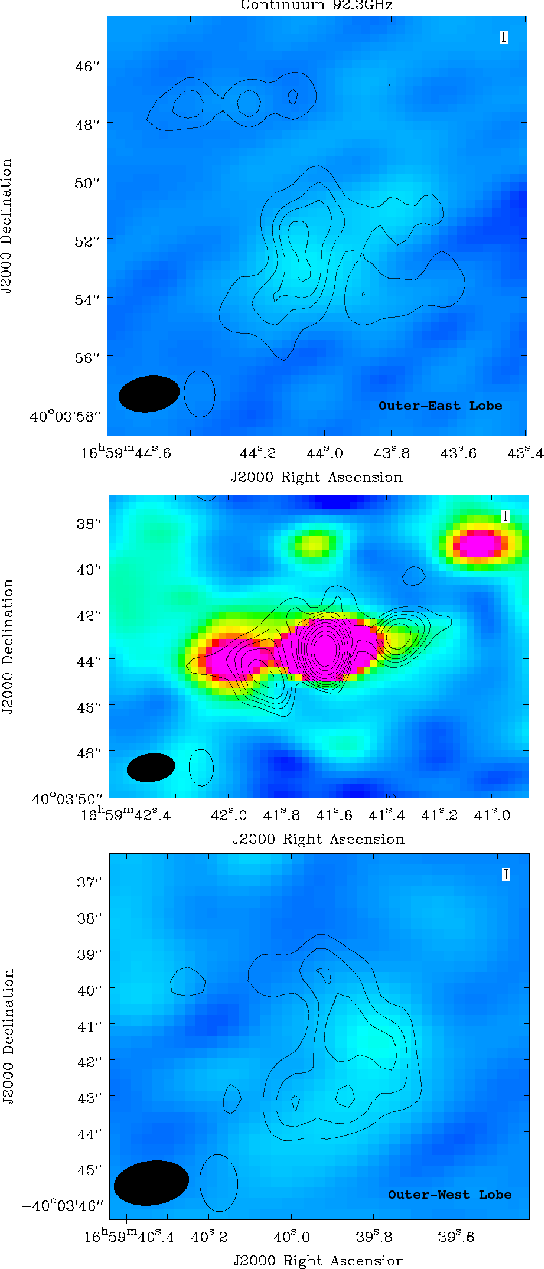}}%
\figcaption{Comparison between the {92 GHz} continuum (color
  background image) and 8.6 GHz emission (black contours) reported by
  \citet{Guzman2010ApJ} {following their naming convention}. In the
  bottom left corner of each panel, we show the synthesized beams,
  filled and outlined for the 92 GHz and 8.6 GHz data, respectively.
  \emph{Top and bottom panel:} Emission detected toward the O-E and
  O-W lobes, respectively.  The central panel shows the emission
  toward the inner lobes flanking the central jet source identified by
  \citet{Guzman2010ApJ}.\label{fig-AlmaAtca}}
\end{figure}

We distinguish three groups of continuum sources associated with the IRAS
16562$-$3959 clump, classified according to their spectral indexes
given in Table \ref{tab-ClumpSou}. The indices  allow us to
propose three plausible mechanisms for the emission.
\begin{enumerate}
\item{Sources with spectral index $\sim1$ (Sources 6, 10, and 18):
  These indexes are characteristic of ionized thermal jets and
  hypercompact \hii\ regions \citep[HC\hii\ regions][]{Guzman2012ApJ,Keto2008ApJ} and indicate partially optically
  thick free-free emission. Source 10 is coincident with the jet
  detected by \citet{Guzman2010ApJ}, which is the main interest of the
  present work and {which} we  analyze in detail in the next sections.  The
  position of Source 18 coincides with the mid-IR source GLIMPSE 
  G345.4977+01.4668 \citep{Benjamin2003PASP} . 
  An exploration of YSO models with mid-IR fluxes
  consistent with Source 18, made using the online fitter
  tool\footnote{http://caravan.astro.wisc.edu/protostars/} described
  in \citet{Robitaille2007ApJS}, indicates that this source is likely
  an intermediate-mass young star with  a  luminosity of 
  $\sim10^4\Lsun$. 
  The   association {with} an OH maser \citep{Caswell1998MNRAS297215} supports
  this interpretation.  The {92 GHz} emission from Source 18 most likely
  arises from a photo-ionized HC\hii\ region or a stellar wind.  This is
  also possibly the case for Source 6, although it is one of the faintest
  sources detected in the field.}
\item{Sources with flat spectrum (Source 1): This spectrum 
  is characteristic of  optically thin free-free emission.  In 
  Source 1, the excitation arises most likely from shocks, since it is 
  associated with  one of the ionized lobes of G345.49+1.47 (see Figure \ref{fig-AlmaAtca}).}
\item{{ Sources with spectral indices $>2$. Fourteen of the eighteen
  sources are in this category. They have a mean spectral index of
  3.2. For the case of isothermal free-free emission, the spectral
  index must remain between $-0.1$ and 2 for all density distributions
  \citep{Rodriguez1993RMxAA}. However, if the ionized medium has a
  temperature gradient, then large spectral indices can be obtained
  \citep{Reynolds1986ApJ}. A more likely explanation is, however, that
  the spectrum near 100 GHz is substantially affected by optically
  thin dust
  \citep[e.g.~][]{Zhang2007ApJ,Beuther2007AA,Zapata2009ApJ,Galvan-Madrid2010ApJ,Maud2013MNRAS}. Since
  the dust emissivity varies as $\nu^\beta$, optically thin emission in
  the Rayleigh-Jeans limit has a spectral index of $\beta + 2$. Values
  of $\beta \sim 1$ can be attributed to grain growth in dense
  environments \citep{Draine2006ApJ}. However, our spectral indices were
  derived over a very narrow frequency range of 85-100 GHz.  Hence,
  since we cannot separate the free-free component from the dust
  component, we do not draw any definite conclusions about the nature
  of the dust grains.  We propose that all these sources are
  associated with dust emission arising from molecular cores within
  the young stellar population of IRAS 16562$-$3959.  Among these, we
  highlight Sources 12 and 13. Source 12 position is consistent with
  IR sources GLIMPSE G345.4906+01.4655 and 2MASS 16594180-4003591
  \citep{Skrustskie2006AJ}, and the results of the YSO online fitting
  tool indicate that it corresponds to a source with a bolometric
  luminosity of $\sim1000\Lsun$. Source 13 is the second most
  luminous source detected toward the IRAS 16562$-$3959 field, and its
  position suggests that it may harbor the powering source of the
  North-South bipolar molecular outflow detected by
  \citet{Guzman2011ApJ}. Finally, we note that Sources 9 and 14 seem to be
  associated with  mid-IR point sources detected by IRAC at 4.5 \micron\ \citep{Fazio2004ApJS,Benjamin2003PASP}.}}
\end{enumerate}

{\subsection{The ionized jet  observed at 92 GHz}\label{sec-jetat3}}

The three panels of Figure \ref{fig-AlmaAtca} show images of the {92 GHz}
emission observed toward the jet and the lobe system reported by
\citet{Guzman2010ApJ}, overlaid with contours at 8.6 GHz emission.
{In this work, we refer to 92 GHz emission as the emission calculated combining the four SpWs.} 
Following the \citet{Guzman2010ApJ} naming convention, the two outermost
lobes are referred to as outer-east (O-E) and outer-west (O-W) lobes,
while the two innermost as inner-east (I-E) and inner-west (I-W)
lobes.

Source 10 coincides, within the uncertainties, with the jet source.  Furthermore,
the spectral indexes of the radio continuum from the jet below 10 GHz
($0.85\pm0.15$) and of the {92 GHz} emission of Source 10 ($1.0\pm0.1$)
are similar.  We conclude that Source 10 is the {92 GHz} counterpart of
the ionized jet and that the emission at
both frequency ranges comes from
partially optically thick ionized gas.

\begin{figure}
\includegraphics[width=\textwidth]{{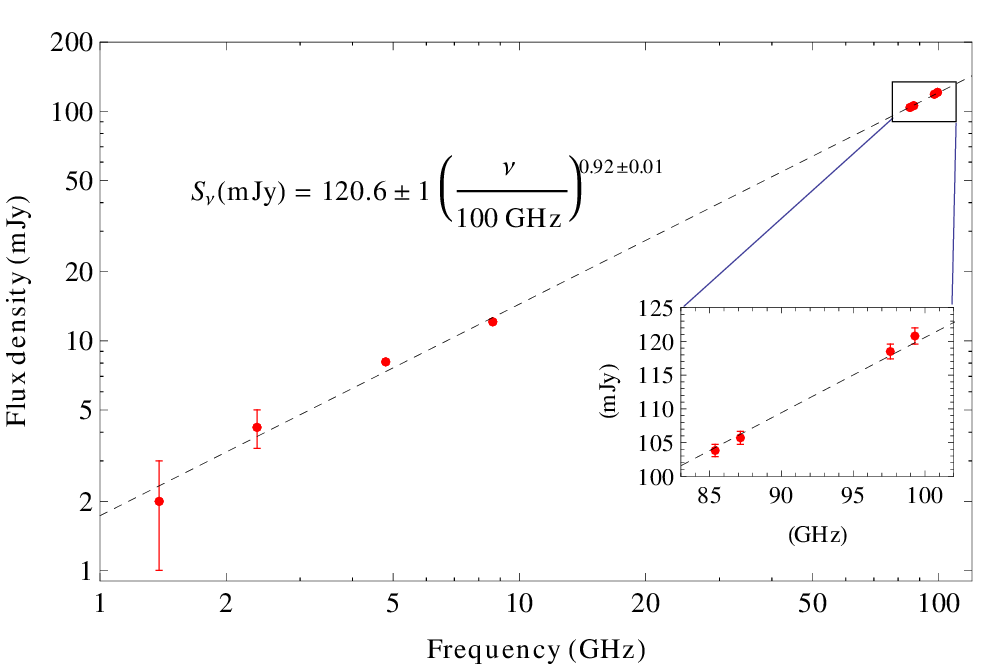}}%
\figcaption{Radio continuum flux density vs.\ frequency (log-log
  plot) for G345.49+1.47.  The dashed line indicates a least-squares
  power law fit over the whole frequency range, giving an spectral
  index of $0.92$.  The fit parameters of the spectrum are shown in
  the figure. The inset displays a zoom over the range of data from
  ALMA.  ATCA data shown for frequencies below 10 GHz are taken from
  \citet{Guzman2010ApJ}.\label{fig-ContSpec}}
\end{figure}

Figure \ref{fig-ContSpec} presents the radio continuum spectra of the
ionized jet in the range from 1 to 100 GHz. This spectra is well
fitted by a power law in frequency with an spectral index of
$0.92\pm0.01$ over the entire frequency range, 
consistent with partially optically thick thermal free-free emission 
\citep{Anglada1996ASPC,Villuendas1996ASPC,Jaffe1999ApJ}.  
While this is the dominating emission mechanism, it is 
likely 
that at the highest frequencies a small 
fraction of the emission arises from thermal dust. We note
that the ALMA spectrum of  Source 10 is marginally steeper
than the one measured {at centimeter wavelengths} alone, which might be
attributed to dust emission. By fitting the data with two power laws,
one for partially optically thick free-free emission and the other
with an spectral index equal to 3, representing the dust contribution,
we find that the flux attributable to dust is
$\sim11$~mJy at 92 GHz, that is, $\sim$10\% of the total flux coming from 
G345.49+1.47.

The ionizing photon flux needed to maintain the recombination
equilibrium is $\ge5.8\times10^{46}$~s$^{-1}$ \citep{Guzman2012ApJ},
which is larger than the typical ionizing fluxes needed to maintain
the typical mJy-level flux of jets detected in centimeter bands.  
The momentum
rate estimated for G345.49+1.47, of  $\sim 10^{-3}$~\Msun~yr$^{-1}$~\kms
\citep{Guzman2010ApJ}, is approximately two orders of magnitude
smaller than the minimum required for the jet in order to shock-ionize
itself \citep{Johnston2013AA,Curiel1989ApLC}. 
Furthermore, as analyzed
in the subsequent sections, the value  of the momentum rate computed for 
G345.49+1.47 is likely overestimated since the velocity of the
ionized gas is $<300$~\kms. 
We conclude that the ionizing UV photons
come from the young high-mass star itself, from which we expect
fluxes larger than $10^{47}$~s$^{-1}$ \citep[see, for example][]{Martins2005AA}.
This is despite the evidence
that G345.49+1.47 appears to be accreting at a high rate, which in
theory should quench the development of an \hii\ region
\citep{Walmsley1995RMxAACS} or decrease substantially the effective
temperature of the young star \citep{Hoare2007dmsf,Hosokawa2009ApJ}.

We also detected emission 
at the positions of the outer lobes of the jet, as shown in
Figure \ref{fig-AlmaAtca}.  
Source 1 of Table \ref{tab-ClumpSou} corresponds to the O-W lobe.  This
source has a flat spectrum, characteristic of optically thin free-free
emission, which is expected since the emission from the ionized lobes
is already optically thin at centimeter wavelengths.  The O-E lobe is
associated with diffuse emission above $5\sigma$, shown in
Figure \ref{fig-AlmaAtca}.  This identification gives confidence that most
features shown in the color map of Figure \ref{fig-AlmaAtca} are real.
The flux density of the outer lobes at 92 GHz, corrected for primary beam
response and integrated over the regions shown in
Figure \ref{fig-AlmaAtca}, are 4.5 and 5.8 mJy for the O-W and O-E
lobes, respectively.  These fluxes are consistent with the ones
reported by \citet{Guzman2010ApJ}, scaled with frequency as
$\nu^{-0.1}$.

Emission from the inner lobes {is}, however,  difficult to disentangle
from our data. This is partly because the angular resolution of the
ALMA data is approximately two times lower than that of {the centimeter wavelength observations}, 
producing an overlap of 
the emission from the inner lobes with that of nearby sources and from
the jet. We expect the inner lobes to have $\sim 4$~mJy each. Source
13, as remarked previously, displays a dustlike emission spectrum, and
it is not a free-free counterpart of the inner-east lobe.

\subsection{Broadening of the HRLs associated with the jet G345.49+1.47} 

Theoretical work predicts that, in addition to thermal and turbulent
broadening, HRLs should be broadened by the linear
Stark effect, namely, the splitting and displacement of the atomic
energy levels by an electric field.  For interstellar ionized regions
and for HRLs in the radio and millimeter regions of the spectrum, the
most important mechanism for Stark broadening is scattering with
electrons under conditions where the impact approximation is valid
\citep{Griem1967ApJ}.  The Stark broadening redistributes the energy
in the line over a frequency interval larger than that produced by the
thermal and turbulent broadening. The predicted line shape is a Voigt
profile, which corresponds to the convolution of a Gaussian component,
produced by the thermal and turbulent motions of the recombining
atoms, and a Lorentzian component, produced by the impacts with
electrons.  Toward the center of the line, the profile is nearly
Gaussian, whereas toward the wings, it is nearly Lorentzian.

The theory of hydrogen Stark broadening was developed after a few
setbacks \citep[see][for a historical account]{Gordon2002ASSL} in which
astronomical observations played a crucial role.  Recently, the theory
was again questioned by observations \citep{Bell2000PASP}, triggering a
debate settled recently by \citet{Alexander2012ApJ}.  Direct detection
of clear-cut Voigt profiles {is an} important confirmation of the
currently accepted broadening theory, and until now, {these profiles have not}
 been unambiguously observed in the astronomical
context.

Most of the reported HRL observations in the literature are
not sensitive enough to detect the wing emission from the Voigt
profiles, {and consequently},  their profiles appear roughly  Gaussian. The presence
of Stark broadening {is} inferred {indirectly}, usually from
observations of two HRLs with different principal quantum
numbers. Because the pressure broadening increases with the transition
quantum number as $n^{7.5}$, the thermal broadening is determined from
the width of the HRL with the smaller $n$ (higher
frequency), and the pressure broadening is estimated from the increase
of the line width of the high $n$ transition with respect to that of the low
$n$, assuming optically thin conditions  \citep{Keto2008ApJ,Sewilo2011ApJS,Galvan-Madrid2012AA}.  In the few cases where
there is a direct wing  detection, 
it is weak
\citep[e.g.,][]{Simpson1973ApSS,Smirnov1984AA,Foster2007ApJ,vonProchazka2010PASP}. 
Apparently, the clearest previous case of pressure broadening is shown in  
carbon recombination lines observed in absorption towards SNR Cas A \citep[see][and references therein]{Stepkin2007MNRAS}. 
 The lack of direct detection
of the wing emission in Voigt profiles from HRLs is likely due  {both to}
their intrinsic low intensities  and to the wide velocity range they
span, requiring for identification sensitive observations and flat or linear instrumental
baselines that just have become available with the new generation of
radio telescopes.

{\subsubsection{Voigt profile fitting}\label{sec-voigt}}

Figure 6 shows the HRL profiles at the peak position. They display evident wing
emission, which we fit using  Voigt functions.  Figure
\ref{fig-HrlsVoigt} also shows the results of Voigt profile fits and
residuals.  The Voigt function is characterized by four parameters:
the value at the peak, the central velocity,  and the  Lorentzian and Gaussian 
widths.  We parameterize the Gaussian width ($\Delta_G$) using the relation
$\Delta_G=0.22\sqrt{T\,\text{(K)}}$~\kms\ \citep[][\S\,2.2.2]{Gordon2002ASSL},
where $T$ is the temperature of the gas that emits the HRL.  

Best-fit parameters are obtained by minimization of the
weighted-squared difference
\begin{equation}
\sum_{\rm HRLs}\sum_{V_l<v_i<V_m} \left(\frac{{\rm HRL}({v_i})-\mathcal{V}({\rm Peak},V_0,\delta_L,T;v_i)}{\sigma}
\right)^2~~,\label{eq-chi2voigt}
\end{equation}
where $\sigma=1$~mJy is the noise per channel, the first sum runs over the three recombination line data
 (${\rm HRL}=$\ha{40}, \ha{42}, \hb{50}),  and the second sum runs over
the channels ($v_i$) between the $V_{\rm LSR}$ limits $V_l=-80$ and
$V_m=100$~\kms. These velocity limits exclude the regions where we
expect contamination from He or C recombination lines, or molecular
lines.  The expression
$\mathcal{V}({\rm Peak},V_0,\delta_L,T;v_i)$ represents the value of
the Voigt function associated with its four parameters
(${\rm Peak}, V_0, \delta_L$ and $T$),  evaluated at velocity $v_i$.
The Voigt function parameters are allowed to be different between the
HRLs, except for $T$.  Our best-fit model indicates that this
temperature is $T=2000^{+6000}_{-1000}$~K. The 1-$\sigma$ uncertainty
range of best-fit temperatures is large since the thermal Gaussian
width is much narrower than the wings. However, the range of values is
within what is expected for ionized regions in the Galaxy. In
particular, we do not find evidence of nonthermal Gaussian
broadening, as   may be expected from turbulence. In the {following}
analysis, we fixed the value of the temperature of the ionized gas as
$T=7000$~K, which is 
close to the expected electronic temperature for the
galactocentric distance of G345.49+1.47 \citep[6400
  K,][]{Paladini2004MNRAS} and within the  uncertainty.
 
Table \ref{tab-Voigt} lists the best-fit values and uncertainties of
the Voigt-function fittings to the three HRLs detected, assuming
$T=7000$~K.  The central velocity of the three lines is the same
within the errors.  The half-width at half-maximum of the Lorentzian
portion of the profiles, $\delta_L$, is between 18 and 20
km~s$^{-1}$.  The Voigt function model, with and without constrained
temperature, performs better than single or double Gaussian models
evaluated by a heuristic visual assessment and according to the
quantitative Akaike information criterion \citep[AIC,][\S\,3.7.3]{Feigelson2012msma}. The AIC penalizes the weighted least-squared
difference by adding two times the number of free parameters of the
specific model. The free parameters of the Voigt model are 10, while for two
independent Gaussians per HRL they are 18. The correlation between adjacent
channels introduced by the Hanning smoothing of the data does not
affect the validity of the application of the AIC.

\begin{deluxetable}{lcccc}
   \tablewidth{0pc} \tablecolumns{5} \tabletypesize{\small}
   \tablecaption{Voigt fitting parameters of the HRLs\label{tab-Voigt}}
   \tablehead{ 
     \colhead{}&\colhead{Peak Flux}&\colhead{$V_0$}&\colhead{${\delta_L}$}&\colhead{Characteristic} \\%
     
     \colhead{}&\colhead{Density}&\colhead{(LSR)}&\colhead{(\kms)}&\colhead{Density}\\%
     \colhead{}&\colhead{(mJy)}& \colhead{(\kms)}  &\colhead{}  & \colhead{($10^7\,$cm$^{-3}$)}}
\startdata
H40$\alpha$    &$41.4\pm0.8$&$-14.8\pm0.5$ & $18.6\pm1.0$  &  $9.7\pm0.5$   \\
H42$\alpha$    &$30.1\pm0.8$&$-14.8\pm0.7$ & $19.2\pm1.0$  &  $6.9\pm0.4$  \\
H50$\beta$     &$7.7\pm0.8$ &$-13.3\pm3.0$ & $20.2\pm4.0$  &  $3.8\pm0.8$
\enddata
\tablecomments{Assuming $T_{e} = 7000$~K for the three lines.}
\end{deluxetable}

\subsubsection{Electron density }

The electron density can be derived theoretically from  Voigt
fitting {of the HRLs} using the relation between $\delta_L$ and the physical
parameters of the ionized gas, first computed by
\citet{Griem1967ApJ}. For the range of principal quantum numbers
appropriate for this work, this relation can be written to an adequate
accuracy level as \citep{Walmsley1990AAS}
\begin{equation}
\delta_L= 2.72~{\rm km~s^{-1}}\left(\frac{n}{42}\right)^{7.5}{\Delta n}^{-1}\left(\frac{N_e}{10^7~{\rm cm^{-3}}}\right)
\left(\frac{T_e}{\rm  10^4~K}\right)^{-0.1}~.\label{eq-density}
\end{equation}  
In Equation \eqref{eq-density}, the HRL is produced by the decay of an
electron from the $(n+\Delta n)$ to the $n$ quantum level. $T_e$ is
the electron temperature, and $N_e$ is the free electron density.  The
last column of Table \ref{tab-Voigt} gives the electron densities
derived from the observed values of $\delta_L$.  We find a
characteristic value for the electron density of $5\times10^7$
cm$^{-3}$.

{\subsection{Ionized wind model: continuum and recombination lines}\label{sec-model}}

In this section, we  present a simple model of the jet
that {explains simultaneously} the  main characteristics of the continuum and HRL
emission.  

{\subsubsection{Collimated ionized jet model and continuum spectrum}\label{sec-modelDesc}}

\begin{figure}
\includegraphics[ width=\textwidth]{{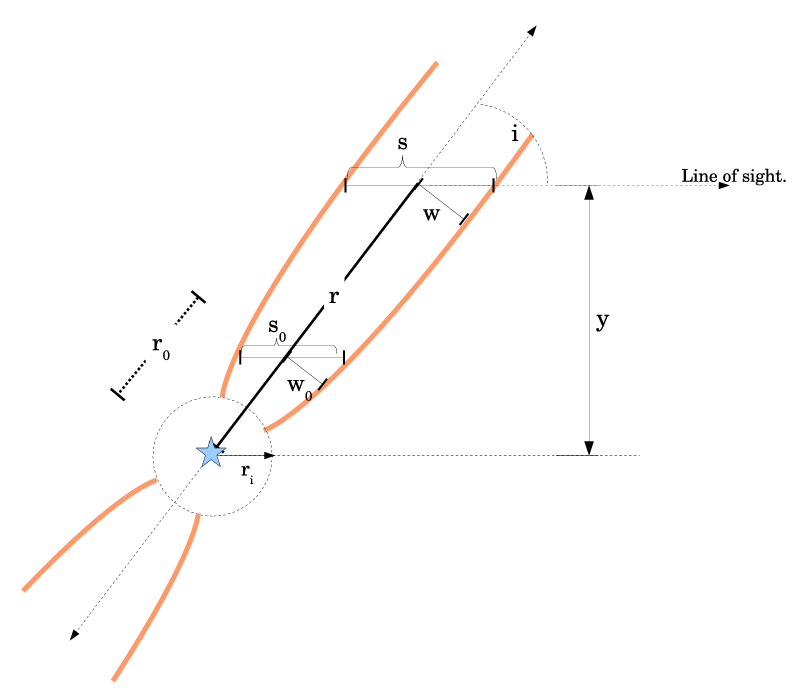}}
\figcaption{Geometrical model of the jet plus a symmetrical
counterjet. This figure is based on Figure 1 from
\citet{Reynolds1986ApJ}.  It represents a cut of the
jet+counterjet system in the plane that includes the axis of the jet
and the line of sight. The jet+counterjet system is
assumed to be axisymmetric. \label{fig-dibujo}}
\end{figure}

A useful parameterization of the jet structure is given in
\citet{Reynolds1986ApJ}. This parameterization is flexible enough to
reproduce the continuum emission spectrum for most ionized jetlike sources,
including G345.49+1.47 \citep{Guzman2010ApJ}.  We use the same notation
as \citet{Reynolds1986ApJ}, with a few differences that are made
explicit below. Figure \ref{fig-dibujo} depicts the {geometry of the
jet model}. Quantities with  a 0-subscript  ($r_0, w_0$, etc.\ldots)
correspond to those at a fixed fiducial radius $r_0$, rather than at
the inner termination radius of the jet, as used  in
\citet{Reynolds1986ApJ}. We choose as the fiducial radius 
$r_0=100$~AU.\footnote{\citet{Reynolds1986ApJ} used $10^{15}$~cm,
  which is 67 AU.}  For  parameters at the 
the inner termination radius, we  use an
i-subscript   ($r_{\rm i}, w_{\rm i}$, etc.\ldots), and for  quantities
at the outer limit of the jet, an f-subscript.

We use  the following relations
\begin{align}
w(r)&=w_0\left(\frac{r}{r_0}\right)^\epsilon&&\text{aperture power law,}\label{eq-epsilon}\\
y(r)&=r \sin(i)&&\text{projected distance,}\label{proj-height}\\
\theta_0&=\frac{2 w_0}{r_0}&&\text{collimation factor,}\\
s(r)&=\frac{2 w(r)}{\sin(i)}&&\text{path length.}
\end{align} 
The shape of the jet is given by $\epsilon$ and $\theta_0$, with
$0.5<\epsilon\le1$. A conical wind corresponds to  $\epsilon=1$.
The physical size of the jet is given by $w_0$, $r_{\rm i}$, and $r_{\rm f}$. 
We assume that the ionized gas consists {only of}
hydrogen, with a constant ionization fraction.  
Therefore, the density
(in cm$^{-3}$) of the ionized gas is equal to two times the free electron
density $N_e(r)$.  
The density $N_e(r)$,  velocity $v(r)$, and
temperature $T(r)$ are assumed to  depend as power laws with  the distance from the
origin as
\begin{align}
N_e(r)&=N_{e,0}(r/r_0)^{q_n}~~,\label{eq-densitylaw}\\
v(r)&=v_0(r/r_0)^{q_v}~~,\label{eq-velocitylaw}\\
T(r)&=T_0(r/r_0)^{q_T}~~.
\end{align} 
If we assume that the velocity of the ionized gas is in the axis
direction and away from the origin, $q_v>0$ and $q_v<0$
represent, respectively, accelerating and decelerating winds.
Considering a constant ionization fraction, mass conservation implies
that $q_n=-q_v-2\epsilon$.  We assume that the free-free absorption
coefficient ($\kappa_\nu$) has  a power law dependence with
temperature ($\propto T^{-1.35}$) and density \citep[$\propto N_e^2$, see, for example,][\S 10.6]{Wilson2009SV}.  This
assumption implies that the free-free optical depth associated with a
line of sight that intersects the jet axis, given approximately by
$\tau(\nu,r)\approx \kappa_{\nu}(r)\times s(r)$, also behaves as a
power law on $r$ given by
\begin{align}
\tau(\nu,r)&=\tau(\nu,0)(r/r_0)^{q_\tau}~~,\label{eq-opacitylaw}\\
\text{with}\ q_\tau&=\epsilon+2 q_n-1.35q_T~~.\label{eq-opindex}
\end{align}

The flux density predicted  from the jet model presented here is derived in
appendix \ref{ap-continuum}. In the intermediate range of frequencies
where the jet is neither completely optically thin {nor} thick, the flux
density is given by
\begin{align}
S_\nu&= S_{\nu_0}\left(\frac{\nu}{\nu_0}\right)^{\alpha_{\rm op}}~~,\label{eq-Snu}\\
\alpha_{\rm op}&=2+2.1(1+\epsilon+q_T)/q_\tau~~,\label{eq-specindex}
\end{align}
where $S_{\nu_0}$ is the flux at the fiducial frequency
$\nu_0$.  Equation \eqref{eq-Snu} is a rising power law in frequency, as
 observed in G345.49+1.47 over almost two decades
in frequency (see \S\,\ref{sec-jetat3}) and toward several other jets and
broad-HRL HC\hii\ regions \citep[see][and references
  therein]{Guzman2012ApJ,Jaffe1999ApJ}.

The most important caveat associated with the model  is that
the data constraints on the parameters are not tight.  The selection of
particular parameters  is heuristic and starts with what is
considered the simplest choice.  In our case, we will explore
 \emph{isothermal} models ($q_T=0$)   with a \emph{constant
  ionization fraction}. These constraints are not sufficient to
determine uniquely the jet parameters. There are still  five
free parameters: the inclination angle ($i$), the fiducial aperture 
($\theta_0$), the density at the fiducial radius
($N_{e,0}$), and two exponents: the shape exponent ($\epsilon$) and
the density exponent ($q_n$). The constraints imposed by the {spectral energy distribution},
given in Equation \eqref{eq-Snu}, are only two: $S_{\nu_0}$ and
$\alpha_{\rm op}$.
The inclination angle $i$ was estimated as $\sim45$\arcdeg\ by
\citet{Guzman2010ApJ},  based on the appearance of the 2~\micron\ image
of the inner cavity, and we  use this value throughout.

With these constraints, we find that the observed continuum spectrum of
G345.49+1.47 is well modeled as free-free emission arising from a
fully ionized, isothermal (7000 K), conical ($\epsilon=1$) wind, with a
collimation factor $\theta_0=0.33$.  There are still
two degrees of freedom on the model, and  this particular
selection of $\epsilon$ and $\theta_0$ is arbitrary.

{\subsubsection{Model prediction for  HRLs}\label{sec-modelHRL}}

In this section, we discuss the HRLs expected from the model presented
in the previous section.  The goal is to test whether we can
reproduce the observed line fluxes and profiles using only Stark and thermal Gaussian
broadening.  This would imply, at least as far our observations can
constrain, that the observed line wings are not due to 
bulk motions of the ionized gas.  
Even though the lines seem to be 
well reproduced by Voigt profiles, {there is a problem with the interpretation of}  the pressure broadening: 
the {observed ratio} between the line widths
of the $\alpha$ and $\beta$ transitions is not as expected from
Equation \eqref{eq-density}, using a single characteristic
density. Equation \eqref{eq-density} predicts that the FWHMs should be
in the ratio $\sim(50/40)^{7.5}/2\approx2.7$, while the observed value
is $\sim1$. We will come back to this issue at the end of the section. 

The line flux expected from the jet model described in the previous
section, $S_{\rm\nu, L}$, {under assumptions of isothermality and  local thermodynamic equilibrium (LTE),} is
derived in appendix \ref{ap-hrls}. $S_{\rm\nu, L}$  is given by
\begin{equation}
%S_{\rm\nu, L}=
S_{\rm\nu_0}\left(\frac{\nu}{\nu_0}\right)^{\alpha_{op}} \left(\Gamma\left(\frac{\alpha_{\rm op}-2}{2.1}\right)^{-1}\int_0^\infty \tau^{\frac{\alpha_{\rm op}-4.1}{2.1}}\left(1-e^{-\tau (1+\Delta\nu_{\rm L}\phi(\nu,N_e))}\right) d\tau-1\right)~,\label{eq-SL}
\end{equation}
where $\phi(\nu,N_e)$ is the line profile ($\int\phi d\nu=1$), and
$\Delta\nu_L$ (defined in  appendix \ref{ap-hrls}) depends only on the specific
transition and on the gas temperature. For the three lines considered
in this work, $\Delta\nu_L$ takes the value
\begin{equation}
\{\Delta\nu_{\rm H40\alpha},\Delta\nu_{\rm H42\alpha},\Delta\nu_{\rm H50\beta}\}=
\{10.8,~7.9,~3.0\}~{\rm MHz}\times\left(\frac{T_e}{10,000{\rm K}}\right)^{-1.25}~~, \label{eq-ew}
\end{equation}
{where the temperature dependence form is appropriate for frequencies 
around 100 GHz \citep{Lang1999AF}.}
The line profile $\phi(\nu,N_e)$ is a Voigt function with density
parameter $N_e$. This density is not constant but depends on the
integration variable (the continuum opacity $\tau$) as
$N_e(\tau)=N_{e,0}(\tau/\tau_0)^{q_n/q_\tau}$. This relation  can be deduced from
Equations \eqref{eq-densitylaw} and \eqref{eq-opacitylaw}.

We test the hypothesis 
that the observed line wings
are reproducible  with negligible bulk motions. 
{We}  assume {in Equation \eqref{eq-ew}} that the line profiles have the same
central frequency instead of being displaced by their corresponding
Doppler shifts and  that the gas velocity  behaves
as $v_0(r/r_0)^{q_v}$ (Equation \ref{eq-velocitylaw}).  Our hypothesis of
negligible bulk {motions} is equivalent to $v_0$ being  small
compared to the line widths.  

\begin{figure}
\centering\includegraphics[height=0.875\textheight]{{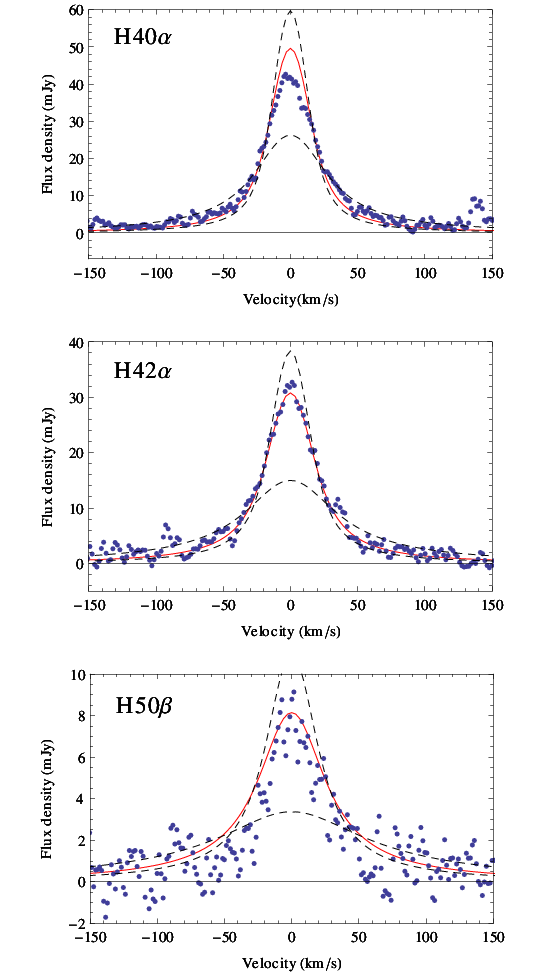}}%
\figcaption{\emph{Blue dots}: Spectra of the observed HRLs toward
  G345.49+1.47.  \emph{Continuous red line}: Prediction of the
  emission arising from a jet model with negligible velocities and
  the parameters given in Equation \eqref{eq-bestparams}. The
  velocity scale is shifted to the common central velocity of the HRLs
  of $-14.8$~km~s$^{-1}$. \emph{Dashed lines}: Model predictions of
  the HRLs using a smaller and larger fiducial aperture compared {with}
  the optimal value, predicting, respectively, the flatter and sharper
  curves. \label{fig-LTE}}
\end{figure}

Figure \ref{fig-LTE} shows with a continuous red line the expected
profiles derived from Equation \eqref{eq-SL} using $S_{\nu_0}=S_{\rm
  100\,GHz}=120.6$ mJy, $\alpha_{op}=0.92$, and $T_0=7000$ K. {Either}  of the
following jet parameters give indistinguishable predictions
\begin{equation}\label{eq-bestparams}
\begin{split}
\epsilon=1&,~~\theta_0=0.33,~~N_{e,0}=2.19\times10^7~\text{cm}^{-3}\quad\text{(conical wind),}\\
\epsilon=0.5&,~~\theta_0=0.23,~~N_{e,0}=2.82\times10^7~\text{cm}^{-3}~~. 
\end{split} 
\end{equation}

For each value of the geometrical aperture
exponent $\epsilon$, there is an optimal  fiducial aperture
$\theta_0$ that predicts approximately the same HRLs.  Therefore,
the HRL observations have decreased the degeneracy degree from 2 (see
previous section) to 1. The optimal $\epsilon$-$\theta_{0}$ curve has
as extreme points the conical ($\epsilon=1$) and $\epsilon=0.5$ aperture laws given
in Equations \eqref{eq-bestparams}. 
The dashed lines 
 in Figure \ref{fig-LTE} show  the 
behavior of the model predictions for $\theta_0$
larger and smaller than the best-fit aperture angle ($\theta_{0,{\rm
    opt}}$). If $\theta_0>\theta_{0,{\rm opt}}$,  a larger
fraction of the gas is more diffuse, increasing the peak-to-width
ratio appearance of the line. This is illustrated in
Figure \ref{fig-LTE} with the narrower and sharp dashed curve,
representing the prediction for a conical wind and $\theta_0=0.4$.  On
the other hand, if $\theta_0<\theta_{0,{\rm opt}}$, more emission
arises from more dense gas, and the HRLs widen, illustrated by the
flatter curve ($\theta_0=0.2$).

We find that the predicted flux level and shape of the lines are
consistent with the data, which is remarkable considering the
simplicity of the model, namely, no bulk motions and LTE
assumptions. Quantitatively, LTE can be justified since the typical
densities derived for G345.49+1.47 ($\sim5\times10^7$~cm$^{-3}$, see
Table \ref{tab-Voigt}) are $\sim1$ order of magnitude larger than the
critical electron density given by Equation (3.1.6) from
\citet{Strelnitski1996ApJ}.  These higher electronic densities increase
the collision rates and damp non-LTE effects.

The model also reproduces the relation observed between the
line widths of the $\alpha$ and $\beta$ transitions, which are not in
the ratio of $2.7$ as {would be} expected from
Equation \eqref{eq-density}. Usually in the literature, a strong dependence
of line width with quantum number is a common criterion to identify
pressure broadening in HRLs from young massive stars (e.g., see
\citealt{Lumsden2012MNRAS}, using IR HRLs; and \citealt{Keto2008ApJ}).
The reason for the unusual behavior of the G345.49+1.47 lines is that
$\alpha$ transitions, in contrast to $\beta$ transitions, are
associated with large line opacities that saturate the line near the
peak, increasing the line width.  The flux-weighted line peak optical
depth for the \hb{50} line is $\sim1$, while for the $\alpha$ lines, it is
$>6$.  Accordingly, we stress that the values deduced for the density
from independent Voigt profiles fittings, given in Table
\ref{tab-Voigt}, represent average values that need to be
interpreted with care.

The agreement between the model and the data is not perfect, showing
discrepancies near the line center of \ha{40} and in the width of the
\hb{50} line.  Typically, the analysis of hydrogen {recombination
  lines} in the millimeter and sub-millimeter wavelengths includes
large corrections due to non-LTE effects
\citep{Mezger1968Sci,Peters2012MNRAS,Baez-Rubio2013AA}.  We might be
observing these effects near the line center where we expect the
largest opacities.  {Finally, we note that despite the model reproduces
a lower ratio between the widths of the $\beta$ and $\alpha$
transitions compared to that associated with optically thin LTE
conditions, the data seems to accent this feature even more.}

\subsection{Sulfuretted molecular emission from G345.49+1.47}
Sulfuretted molecules, and specially sulfur oxides, seem to increase
their abundance $\sim$3-4 orders of magnitude in high-mass star formation regions
when evolving from IR luminous massive cores \citep{Herpin2009AA} to a
hot core phase \citep{vanderTak2003AA,Jimenez-Serra2012ApJ}.  These
molecules, particularly SO$_2$, have become common tracers of rotation
in disklike structures associated with HMYSOs (see
\citealt{Fernandez-Lopez2011AJ} and references therein; also
\citealt{Jimenez-Serra2012ApJ}).  Figures \ref{fig-SO100m0m1} and
\ref{fig-OxoSrot} show that this also seems to be the case for
G345.49+1.47.

The analysis presented in the
following sections has made extensive use of the Splatalogue,
  JPL, CDMS and Basecol\footnote{http://basecol.obspm.fr} \citep{Dubernet2013AA}
catalogs and databases to obtain transition frequencies, energies,
partition functions, Einstein coefficients, and collision rate coefficients.

\subsubsection{Qualitative chemical analysis of the sulfuretted molecules emission distribution}

All sulfuretted molecules in Table \ref{tab-Slines} display compact
emission associated with the central source G345.49+1.47. While the
sulfur oxides emission  coincides with the jet location and several of
their transitions show characteristic disklike velocity gradients,
the OCS and especially the CS and C$^{33}$S 
lines peak away from the jet and do not exhibit velocity gradients.
The morphological similarity of the CS and C$^{33}$S maps 
suggests  that, at least for CS, 
self-absorption is not the
main cause of the observed displacement.
Based on  Earth sulfur isotopic ratios
$[\,^{32}{\rm S}:\,^{34}{\rm S}:\,^{33}{\rm S}]=[126.7:5.6:1.0]$
given by \citet{DeBievre1993IJMSIP}, we expect an opacity of the
C$^{33}$S line   $\sim130$ times smaller than that of the main
isotopologue. 

In the rest of this section, we address the following questions: 
i) What determines which transitions trace
the velocity gradient that we attribute to a disklike structure?
ii) Why do sulfur oxides seem to be  intimately associated with the
G345.49+1.47 jet free-free continuum?  ii) Why do the OCS and CS lines peak
away from the jet location? 

To answer the first question, we note that  two groups of
lines  do not show the disklike velocity gradient:  the SO$_2$
transitions associated with upper energy levels 
$>200$~K, and  the OCS and CS transitions.  We attribute the lack
of detection of the velocity gradient in the first case to the
relatively poor angular resolution of our data. 
The emission from the 
high-energy SO$_2$ lines most likely {arises} from the hot inner regions of the disklike
structure, located close to the central young star (or stars).  
If this is the case, we should  detect the velocity gradients
in these transitions using better angular resolution observations.  
On the other
hand, emission from the CS and
C$^{33}$S lines {arises}  from a different 
location  compared to  the 
sulfur oxide lines.  

{The displacement between the OCS and sulfur oxides emission  
is smaller, but it is highly unlikely the  OCS emission
is associated with an unresolved hot gas component because,  as derived in section
\ref{sec-Tex}, OCS is in a relatively low excitation state
($\sim40$~K).}  Subthermal excitation
does not play a  role, since the density estimation made from
dust continuum in \citet{Guzman2010ApJ} ($\gtrsim10^5$~cm$^{-3}$) is
above the OCS transitions' critical densities of approximately
$3\times10^4$~cm$^{-3}$ \citep{Green1978ApJ}.  Summarizing, we propose
that the high-energy SO$_2$ lines trace an unresolved hot component
and hence do not show a velocity gradient, and that OCS and CS arise
from the outer gas near the G345.49+1.47 core.

The {close} match of the SO and SO$_2$ emission with the central
free-free jet emission is consistent with the hot core model of
\citet{Charnley1997ApJ}. According to this model, SO and SO$_2$ are created
in the gas phase on timescales $\sim10^{3\text{-}4}$~yr, while OCS and CS
arise in $\sim10^{4\text{-}5}$~yr.  Therefore, since we expect G345.49+1.47
to be younger than $10^5$~yr \citep{Guzman2012thesis}, SO, $^{34}$SO,
and SO$_2$ have been synthesized in the irradiated disklike
structure, but only insignificant amounts of OCS or CS would have been formed. 
 This explains the \emph{absence} of
OCS and CS in the rotating disklike structure, but, why {does
 OCS appear to be}  associated with the G345.49+1.47 core?
\citet{Charnley1997ApJ}, \citet{Hatchell1998AA}, and
\citet{vanderTak2003AA}, all report difficulties in reproducing
the observed abundance  of OCS from observed hot cores,
resorting to grain-mantle chemistry and evaporation from solid-phase
ices as additional sources of OCS.  SO$_2$ and OCS have been detected
in interstellar ices \citep{Gibb2004APJS}, so we adhere to this as a
plausible possibility. We propose that the OCS emission originated
from evaporated ices near the G345.49+1.47 core.

If evaporated ices are indeed the source of the gas-phase OCS, we have
to ask why OCS is absent from the rotating disklike structure.  Why is
there no ice-evaporated OCS associated with the rotating core?  The
study made by \citet{Ferrante2008ApJ} {may} provide an answer: they
report that, under laboratory conditions, high-energy
irradiation\footnote{\citet{Ferrante2008ApJ} uses proton irradiation,
  but photo- and radiation chemical processing of ices is very similar
  \citep{Hudson2000AA}.} of ices synthesizes OCS in the solid phase,
but it is easily destroyed by prolonged {radiation} exposure. It is
possible then that the UV-exposed disk ices are depleted {of} OCS.
This possibility is also consistent with the absence of CS in the
disklike structure, since CS is not produced as result of the OCS
destruction \citep{Ferrante2008ApJ}. In fact, CS does not appear to be
formed within sulfur-containing ices \citep{Maity2013ApJ}.

At least qualitatively, there seems to be a consistent theoretical
picture that explains the presence of sulfur oxides in the directly
irradiated disklike molecular structure and, at the same time,
explains the OCS distribution. We also note that the CS and OCS spatial
distributions are not the same: they peak at different locations, and
the fraction of spread {CS emission is larger compared with OCS}.  
In general, our results are compatible with the
results of \citet{vanderTak2003AA} and \citet{Wakelam2011AA} that find
that CS traces the chemically inactive envelope surrounding the HMYSOs
and hot cores.  It seems now  clear that the strong \cs{2}{1} emission
detected toward IRAS 16562$-$3959, first reported by
\citet{Bronfman1996AAS} using single-dish observations, traces the dense molecular
gas on a clump scale with a limited contribution coming from the
compact central core.

{\subsubsection{Excitation temperatures}\label{sec-Tex}} 

Deriving physical parameters from the observed molecular emission from
a model assuming single excitation temperature (SET) {conditions} 
\citep{vanderTak2011IAUS,Guzman2012thesis} gives some physical insight
into the conditions of the gas in G345.49+1.47.

We briefly describe the main relations and hypotheses behind the
modeling of molecular lines. Detailed  discussions
are given in \citet{Garden1991ApJ}, \citet{Sanhueza2012ApJ}, and
\citet{Wilson2009SV}.  Assuming optically thin and SET conditions,
\begin{equation}
\begin{split}\label{eq-W}
W=\int\int I_\nu d\Omega dv&=B_\nu(T)\Omega_s\int\tau_\nu\,dv\\
                          &=\frac{h c}{4 \pi}\Omega_s A_{\rm ul}N_{\rm u}~~,
\end{split}
\end{equation}
where $W$ is the velocity-integrated line-flux density,
$B_\nu(T)$ is the Planck function evaluated at the excitation
temperature, $\tau_\nu$ is the line opacity, $\Omega_s$ is the solid
angle of the source, $A_{\rm ul}$ is the Einstein A-coefficient of the
transition, and $N_{\rm u}$ is the column density of the molecules in
the upper level of the transition. The integration in velocity covers
the spectral extent of the line.  We define the total luminosity of the
line $\mathcal{L}:=4\pi d^2 W$, where $d=1.7$~kpc.  The
relationship between the population in the upper state and the
temperature is given by the Boltzmann equation,
\begin{equation}
\frac{N_u}{g_u}=\frac{N_X}{Q_X(T)}\exp\left(-\frac{E_u}{k T}\right)~~,\label{eq-be}
\end{equation} 
where $N_X$ is the total column density of species $X$ (SO$_2$, OCS,
etc.\ldots), $g_u$ is the statistical weight of the upper level, $E_u$
{is} its energy (column 4 of Table \ref{tab-Slines}) and $Q_X(T)$
is the partition function evaluated at temperature $T$.  Analogous to
$\mathcal{L}$, we define $\mathcal{N}_X:=d^2 \Omega_s N_X$, the
total number of $X$-molecules in the source.
The critical density of the SO transitions is approximately
$2.8\times10^{5}$~cm$^{-3}$.  We did not find rate coefficients for
the high-energy SO$_2$ transitions, so we assume them equal to
$10^{-11}$~cm$^{3}$~s$^{-1}$. With this assumption, the critical
density is $2\times10^{5}$~cm$^{-3}$, close to that of the SO
transitions. These critical densities are similar to the density
estimation for the inner parts of the IRAS 16562$-$3959 clump
\citep{Guzman2010ApJ}. Therefore, we do not expect that subthermal
excitation has an important observable effect on our data.

We combine Equations \eqref{eq-W} and \eqref{eq-be} and obtain
\begin{equation} 
\frac{\mathcal{L}_{X}}{h c A_{\rm u,l}g_u}= 
\frac{\mathcal{N}_X}{Q_X(T)}\exp\left(-\frac{E_u}{k T}\right)~~,\label{eq-diagram}
\end{equation}
where the left side  of the equation has observable
quantities and the right side has two free parameters per molecular species
($\mathcal{N}_X$ and $T$).

\begin{figure}
\includegraphics[width=\textwidth]{{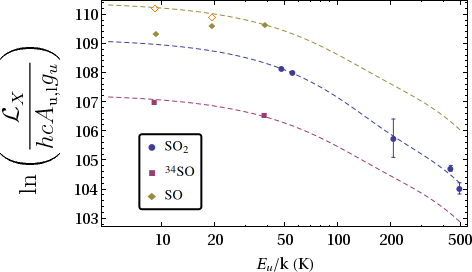}}%
\figcaption{Boltzmann plot of the SO, $^{34}$SO, and SO$_2$ molecular
  line emission associated with G345.49+1.47. Dashed lines represent
  fits using two independent thin-SET models, with the same two
  temperatures fitted to the $^{34}$SO and SO$_2$ transitions. The
  model for SO was derived from the $^{34}$SO {data,} assuming the
  same temperatures and the isotopic ratio
  [$^{32}$S/$^{34}$S]$=22.5$. Empty diamond symbols are the
  opacity-corrected SO values.\label{fig-bp}}
\end{figure}

Figure \ref{fig-bp} shows the quantity corresponding to the left side of
Equation \eqref{eq-diagram} vs.\ the upper energy of the transition for
the three molecules that display the velocity gradient interpreted as
rotation.  The velocity-integrated line fluxes are taken from column
(6) of Table \ref{tab-Slines}, with typical uncertainties of 15
mJy~\kms.  We find that a single SET model cannot fit simultaneously
the data of the low ($\lesssim50$~K) and high ($\gtrsim200$~K) upper-energy 
molecular transitions.  This is somewhat expected, since the
high-energy SO$_2$ transitions trace the rotating core from locations
closer to the HMYSO compared with the low-energy transitions.  However, 
the addition of two independent SET models can reproduce well the
emission of SO$_2$ and $^{34}$SO. We fit a warm and a hot component,
with temperatures of $140^{+60}_{-20}$ and $35^{+25}_{-20}$ K,
respectively.  Dashed lines in Figure \ref{fig-bp} show the prediction
of the model.

Departures from optically thin predictions are expected for lines whose
opacity is greater than 1, and this may be {so} for the strong
SO transitions.  We evaluate this possibility by comparing the
transitions $3_2\rightarrow2_1$ and $4_5\rightarrow4_4$ of $^{34}$SO
and $^{32}$SO.  {For} two isotopologues under SET
conditions, the integrated line quotient between two matching
transitions is given by
\begin{equation}\frac{W_{1}}{W_{2}}=\frac{1-e^{-\tau_1}}{1-e^{-\tau_2}}~~,\label{eq-quot}\end{equation}
where the optical depths are averaged in the line and the ratio
$\tau_1/\tau_2$ is approximately equal to the abundance ratio between the
isotopologues.  The right side  of Equation \eqref{eq-quot} approaches the
opacity ratio under optically thin conditions and approaches 1 in the
optically thick limit \citep[\S\,3.3.1]{Guzman2012thesis}. We assume
that the abundance ratio between the $^{32}$SO and $^{34}$SO
isotopologues is equal to the terrestrial abundance ratio of the
sulfur isotopes, that is, [$^{32}$S/$^{34}$S]$=22.5$.  The line
ratio of the $3_2\rightarrow2_1$ and $4_5\rightarrow4_4$ transition and derived $^{32}$SO opacities are
\begin{equation}
\begin{split}\label{eq-Sratio}
  W_{\rm ^{32}SO}/W_{\rm ^{34}SO}~~3_2\rightarrow2_1 &= 10.9 \Longrightarrow  \tau_{32}= 1.8 \\ 
  W_{\rm ^{32}SO}/W_{\rm ^{34}SO}~~4_5\rightarrow4_4 &= 24.1 \Longrightarrow  \tau_{32} \ll 1~~.
\end{split}
\end{equation} 

A simple opacity correction can be applied to the optically thin model
by multiplying  the right side of
Equation \eqref{eq-diagram} by $(1-\exp{(-\tau)})/\tau$, where $\tau$ is
the line's optical depth  \citep{Goldsmith1999ApJ}.  We estimate the
opacities of the SO $3_2\rightarrow2_1$ and $4_5\rightarrow4_4$ lines
from Equation \eqref{eq-Sratio}, and that of $2_2\rightarrow1_1$ line from
the SET model assuming the isotopic abundance {ratio}
[$^{32}$S/$^{34}$S]$=22.5$. The yellow dashed line in
Figure \ref{fig-bp} displays the prediction of the right side of
Equation \eqref{eq-diagram} for SO.  Empty diamonds in Figure \ref{fig-bp}
{indicate} the opacity-corrected parameters of the SO lines, {which we} 
find  consistent with the
SET model under the assumed isotopic abundance ratio.  As derived
in Equation \eqref{eq-Sratio}, no correction is associated with the
$4_5\rightarrow4_4$ optically thin transition. We emphasize
that the SO lines were not used to derive the SET model (dashed lines).

To conclude, we mention that the temperature derived from the two OCS
transitions is $\sim40$~K. It is similar to the warm component
temperature determined for the sulfur oxides, but as remarked in the
previous section, OCS does not trace the disklike
rotating structure.  Apparently, at the physical scales probed by our
observations (3000 AU, see next section), both a rotating and a presumably larger
non-rotating envelope coexist.

{\subsubsection{Dynamics of the molecular emission}\label{sec-dynM}}

\begin{figure}
\includegraphics[width=\textwidth]{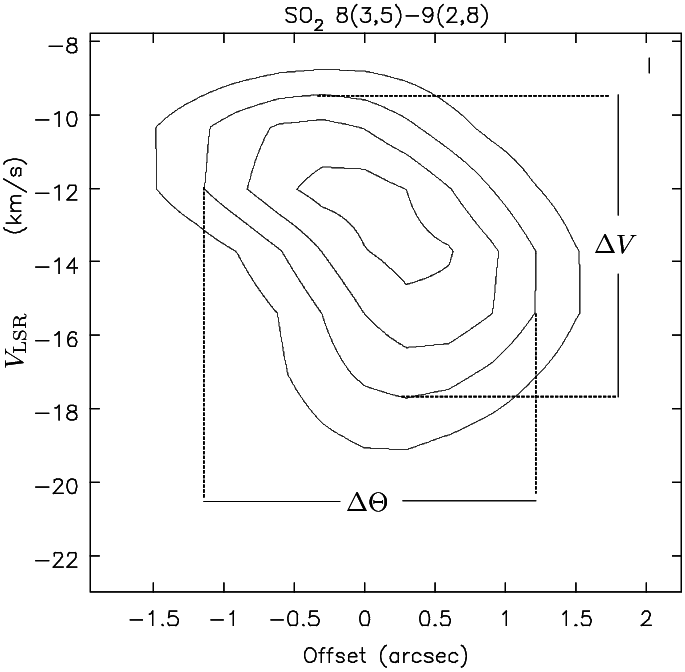}%
\figcaption{Contour map of the position velocity diagram of the SO$_2$
  $8_{3,5}\rightarrow 9_{2,8}$ emission taken across the direction
  with P.A.$=8.9$\arcdeg, through the jet source position (Source 10), averaging
  1\arcsec\ width. The contours correspond to 30, 50, 70, and 90\% of
  the peak equal to 79.2~mJy~beam$^{-1}$. $\Delta V$ and $\Delta
  \Theta $ correspond to the extension of the 50\% contour in velocity
  and angular size, respectively.\label{fig-pv}}
\end{figure}

Figure \ref{fig-pv} shows the position-velocity (PV) diagram measured
from the SO$_2$ $8_{3,5}\rightarrow 9_{2,8}$ emission. The direction of the
PV line is perpendicular to the line indicated in the top left panel
of Figure \ref{fig-OxoSrot} --- the jet direction --- with  zero
offset at the position of the jet source.  This direction is
consistent, within our angular and spectral resolution, with  the
direction of the largest velocity gradient.  It supports the
interpretation that the molecular structure probed by our observations
is part of a rotating  structure with angular momentum
direction aligned with the jet axis.

In order to estimate a dynamical mass, we assume that the disklike structure is
centrifugally supported against the gravity of a central mass.  We estimate the
dynamical mass from the following simplified version of Equation (1) from
\citet{Franco-Hernandez2009ApJ}
\begin{equation}
M_{\star,{\rm dyn}}=\frac{(\sqrt{\Delta\Theta^2-\theta_b^2})\,d}{2 G \sin^2(i)} \left(\frac{\Delta V}{2}\right)^2
~~,\label{eq-dm}
\end{equation}
where $\Delta\Theta$ is the source size, $\Delta V$ is the velocity
breadth, $d$ is the distance, $i$ is the inclination of the disk axis {with}
respect to the line of sight, and $\theta_b$ is the beam size.  From
the 50\% contour of the PV diagram shown in Figure \ref{fig-pv}, we
estimate $\Delta\Theta=2\farcs5$ and $\Delta V=8~\kms$. Using a
distance of 1.7 kpc, $\theta_b= 1\farcs7$, and an inclination of
45\arcdeg\ \citep{Guzman2010ApJ}, we derive a dynamical mass of
$56\Msun$. The approximate physical size of the rotating core, given by
$\sqrt{\Delta\Theta^2-\theta_b^2}\,d$, is $3000$~AU.
{We obtain the same results using instead 
the SO$_2$ $7_{3,5}\rightarrow 8_{2,6}$ transition.}

\subsection{Gentle photo-ionized wind and rotating molecular core toward the HMYSO G345.49+1.47}

Our HRL observations indicate that the ionized gas toward G345.49+1.47
is not moving at a very high velocity ($\sim500$~km~s$^{-1}$), as
observed toward similar objects such as the Cepheus A HW2 jet
\citep{Jimenez-Serra2011ApJ}.  
%%%%%%%%%%%%%%%%
As shown in \S \ref{sec-voigt}, Voigt profiles fit the data adequately
and they relate naturally with the simple physical model presented in
\S \ref{sec-modelHRL}.  In principle, however, the HRLs' wing emission
could be produced by high-velocity outflowing gas analogous to the way
molecular line wings trace molecular outflows.  Line-wing models of
outflows are sufficiently flexible to allow for any decay exponents
between $-1$ and $-4$
\citep[e.g.,][]{Masson1992ApJ,Downes2003AA,Smith1997AA}, and molecular
outflow observations find a distribution of decay exponents also
covering this range \citep{Richer2000prpl}.  
It would be fortuitous, \chg{however}, if
the ionized wings produced by entraining ambient gas had a
Lorentz-like and a symmetric shape. \chg{Furthermore,
between the red- and blue-shifted wing emission 
we detect no shift in the peak position 
larger than 0\farcs2, an upper boundary limited by
our angular resolution (see \S \ref{sec-HRL}).}

Four symmetrically aligned ionized lobes seem to be associated
with the ionized wind of G345.49+1.47 \citep{Guzman2010ApJ}. If these
radio lobes trace shocked-ionized gas,  they must trace high-velocity
shocks.  Proper motion studies have confirmed that radio-lobes
associated with objects similar to G345.49+1.47 move rapidly, determining
velocities close to
$500$~\kms\ \citep{Marti1998ApJ,Curiel2006ApJ,Rodriguez2008AJ}.
Accordingly, it has been often assumed in the literature
\citep[e.g.][]{Garay2003ApJ,Su2004ApJ,Bronfman2008ApJ,Guzman2010ApJ,Guzman2011ApJ,Carrasco-Gonzalez2012ApJ,Johnston2013AA}
that the continuum ionized source detected toward the center of such
systems is tracing high-velocity ionized gas,  and even when 
no lobes are apparent.
 In view of the results presented in this work
toward G345.49+1.47, these assumptions do not seem to be justified unless
supported by complementary HRL data.

The dynamical mass determined in \S\,\ref{sec-dynM} ($56\Msun$) is
larger  than {that of a single high-mass  star 
producing the total luminosity
of IRAS 16562$-$3959 ($70,000\Lsun$, $25\Msun$) or the 
$15\Msun$ estimated for the dominant 
HMYSO}.  
Could the rest of the mass, $30\text{-}40\Msun$ or more, be in the molecular gas phase?  
This is not likely for two reasons. First, most
of the 92 GHz emission comes from ionized gas, with at most $\sim11$~mJy
attributable to dust emission. This is justified in \S\,\ref{sec-jetat3}, 
but also by the HRLs' line intensities, which are
consistent with the continuum.  
This flux  corresponds to only 4
\Msun\ assuming 50 K (ref.~\S\,\ref{sec-Tex}), a dust absorption
coefficient of $0.3$~cm$^{2}$~g$^{-1}$ \citep[extrapolated from the
  coagulated dust tables of][]{Ormel2011AA}, a gas-to-dust mass ratio of 100, and optically thin
conditions. 
Second, comparing the total number of SO$_2$ molecules in
the core, {more than} $30\Msun$ 
 of molecular gas imply an [SO$_2$/H] abundance 
$<6\times10^{-9}$, which seems low compared with the results of
\citet{Wakelam2011AA}. We concede, however, that the SO$_2$ is expected
to vary greatly and this  is not an stringent constraint. We conclude
that it is more probable that there are {(likely more than one)} 
 protostellar companions together with 
the central star. Other compact
components will need higher angular resolution studies to be resolved.

Even though there is no evidence from the observed HRLs {to suggest}
 that the ionized gas is moving, it  is unlikely to be 
 in hydrostatic equilibrium. A  coherent model
for G345.49+1.47 is that of a pressure-accelerated photo-ionized wind,
whose main characteristics can be approximated by a transonic Parker
wind \citep[\S\,3.1.3]{Lamers1999CUP} within a conical aperture
\citep{Lugo2004ApJ,Keto2007ApJ}.  In the Parker wind, the
velocity of gas increases very slowly with distance, following
$(v/a)\approx\sqrt{\ln(r/r_c)}$, with $r_c=G M_\star/2a^2$ and $a=9.8~\kms$ is the isothermal sound speed of a
solar composition ionized gas at 7000 K.  The $\epsilon=1$ {conical} model
presented in \S\,\ref{sec-model} is a rough approximation to this
solution, since it is conical and has the most shallow acceleration
law compared to models with a different geometry ($\epsilon<1$). The
acceleration exponent (Equation \ref{eq-velocitylaw}) takes the form
$q_v=0.97-0.53\epsilon$, as derived from Equation \eqref{eq-specindex} and
mass conservation.  It remains to be seen what would be the  
effects of the inclusion of low wind velocities in the line radiation
transfer.

The wind collimation derived from the model is $\sim3$ (\S\,\ref{sec-modelHRL}), which is not  high
 but nevertheless {is}
comparable to the collimation derived from deconvolved resolved radio
sources believed to be thermal jets, such as NGC 7538 IRS1
\citep{Sandell2009ApJ}, AFGL 2591 \citep{Johnston2013AA}, Cep A HW2
\citep{Curiel2006ApJ}, and G343.1262$-$00.0620
\citep{Rodriguez2005ApJ}.  In addition, the presence of aligned radio
lobes in some cases allows us to infer a more collimated wind such as
in G343.1262$-$00.0620 and G345.49+1.47.  However, what
appears to be clear, at least for G345.49+1.47, 
is that the hypothetical highly collimated fast jet that
excites the ionized lobes does not correspond to the central radio
continuum source.  For the moment, there are no proper motion
measurements toward the lobes of G345.49+1.47, but if they are
rapidly moving ($>300$~\kms, as in G343.1262$-$00.0620, Cep A, and HH
80-81), it would confirm that the ``Jet'' source of
\citet{Guzman2010ApJ} and the lobes are not linked in the way
previously thought.
 
An interesting possibility is that the ionized wind is analogous to
the wide angle low-velocity component observed toward low-mass protostellar
jets \citep[][and references therein]{Torrelles2011MNRAS}.  It is
possible that inside this slow ionized wind exists a much narrower,
denser, and faster jet, powered by accretion, and responsible for the
excitation of the radio lobes. The analogy should not be taken very
far, however, since low-mass stars do not produce photo-ionized
winds.

\section{SUMMARY}
We made observations at frequencies 85-99 GHz using ALMA 
of the continuum, HRLs, and sulfuretted molecular lines 
toward the massive molecular clump IRAS
16562$-$3959, which  harbors the HMYSO G345.49+1.47. 
The main results are summarized as follows:
\begin{enumerate}
\item{We detect \chg{spatially unresolved} emission in the \ha{40},
  \ha{42}, and \hb{50} HRLs toward the collimated ionized wind source
  associated with G345.49+1.47.  The lines display Voigt profiles with
  Lorentzian wings of widths between 30 and 40~\kms, which we
  interpret as pressure broadening arising from ionized gas with
  average density of $5\times10^7$~cm$^{-3}$.}
\item{\chg{A parameterized model of a slow ionized wind is sufficient
    to simultaneously fit the HRLs and the continuum emission between
    1 and 100 GHz associated with G345.49+1.47. There is no need for
    ionized gas moving at velocities in excess of 50~\kms in order to
    explain the HRL profiles.}}
\item{We detect in the ALMA field of view ($\sim 1\arcmin$) at least
  15 additional continuum sources, probably associated with the IRAS
  16562$-$3959 clump, with spectra consistent with {part of the
    emission arising from} optically thin dust. These sources are
  likely to correspond to dusty molecular cores.}
\item{The emission in the SO$_2$, $^{34}$SO, and SO lines with upper energy levels
  $\le50$~K {exhibits} velocity gradients that we interpret as arising from 
  a rotating compact ($\sim3000$~AU) molecular core
  with angular momentum aligned with the jet axis. The estimated 
  dynamical mass is $56\Msun$.}
\item{Sulfuretted molecular emission associated with the core has
  excitation temperatures that range between 35 and 140 K.}
\item{Qualitatively, the SO, SO$_2$, and CS emission and morphology can
  be understood using the predictions of hot gaseous phase chemical
  models (e.g., \citealt{Charnley1997ApJ} and
  \citealt{vanderTak2003AA}).  Additional irradiated ice-chemistry
  might be necessary to explain the characteristics of the OCS
  emission.}
\item{G345.49+1.47 is a $\sim15\Msun$ HMYSO associated with a
  photo-ionized wind that dominates the free-free emission.  It is
  likely that within this photo-ionized wind,  a highly
  collimated jet is powered by an accretion disk, responsible for the
  excitation of the aligned radio lobes.}
\end{enumerate} 

\acknowledgements{The authors are grateful to C. Barrett,
  Y. Contreras, E. Keto, L. Kristensen, Q. Zhang for useful
  discussions and proofreading the manuscript. The authors thank an
  anonymous referee for careful reading and useful suggestions that
  improved this article.  A.E.G.\ acknowledges support from NASA
  Grants NNX12AI55G and NNX10AD68G.\ L.B., G.G., and D.M.\ acknowledge
  support from CONICYT through project PFB-06.  This paper makes use
  of ALMA data {\rm ADS/JAO.ALMA\#2011.0.00351.S}.  ALMA is a
  partnership of ESO (member states), NINS (Japan), NSF (USA), NRC
  (Canada), and NSC and ASIAA (Taiwan), in cooperation with the
  Republic of Chile.  }
%%%%%%%%%%%%%%%%%%%%%%%%%%%%%%%%%%%%%%%%%%%%%%%%%%%%%%%%%%%%%%%%%%%%%%
%% Appendices
%%%%%%%%%%%%%%%%%%%%%%%%%%%%%%%%%%%%%%%%%%%%%%%%%%%%%%%%%%%%%%%%%%%%%%
\appendix
{\section{Continuum flux expected from the jet model}\label{ap-continuum}}
Assuming that the Rayleigh-Jeans approximation is valid at the frequencies 
of interest ($B_\nu(T)\approx2 k T\nu^2/c^2$) and that all relevant quantities 
along a line of sight through the jet are given by their values at the jet 
axis (see Figure \ref{fig-dibujo}),  the radio continuum flux density from the whole jet 
system (jet+counterjet) can be written approximately as (see Equations 7 and 8 
from \citealt{Reynolds1986ApJ})
\begin{align*}
S_\nu &= 2\times\int_{y_{\rm i}}^{y_{\rm f}} \frac{2 k T(y) \nu^2}{c^2} \frac{2 w(y)}{d^2} \left(1-\exp(-\tau(\nu,y))\right) dy\\
&= 2\times \frac{2 k T_0 \nu^2}{c^2} \frac{2 w_0 y_0}{d^2}\int_{y_{\rm i}}^{y_{\rm f}} \left(\frac{y}{y_0}\right)^{\epsilon+q_T}%
\left(1-\exp\left(-\tau_0(\nu)\,(y/y0)^{q_\tau}\right)\right) \frac{dy}{y_0}~~.
\end{align*}
The extra factor  2 with respect to the equations in \citet{Reynolds1986ApJ}
takes into account the emission from the two sides of the jet.
Note that $(r/r_0)=(y/y_0)$, making {it} trivial to change the dependence on $r$ to 
a dependence on $y$. Defining $\Omega_0=2 w_0y_0/d^2$ and 
making the change of variable $u=y/y_0$, we obtain
\begin{equation}
S_\nu =  2\times B_\nu(T_0)\Omega_0 \int_{u_{\rm i}}^{u_{\rm f}} u^{\epsilon+q_T}%
\left(1-\exp\left(-\tau_0(\nu)\,u^{q_\tau}\right)\right) du~~, \label{Snu-intermedio}
\end{equation}
which, after the following change of variable, $\tau=\tau_0(\nu)u^{q_\tau}$ (note that $q_\tau<0$), becomes 
\begin{equation} 
S_\nu=2\times B_\nu(T_0)\Omega_0 \left(\tau_0(\nu)\right)^{-\frac{\epsilon+q_T+1}{q_\tau}}\int_{\tau_{\rm f}}^{\tau_{\rm i}} \tau^{\frac{\epsilon+q_T+1}{q_\tau}-1}\left(1-\exp(-\tau)\right) \frac{d\tau}{(-q_\tau)}~~.\label{Snu-cvtau}
\end{equation}
In this way, the integral over the source extension in the sky is
re-written as an integral over the values taken by the continuum
opacity.  Equation \eqref{Snu-cvtau} gives us the flux density of the
source, which is the only quantity  we can actually probe since we
do not resolve the jet structure.

Defining $\eta:=(1+\epsilon+q_T)/q_\tau$ (called ``$c$'' in \citealt{Reynolds1986ApJ}),
we finally obtain 
\begin{equation}
S_\nu= 2 \times B_\nu(T_0)\Omega_0 \frac{(\tau_0)^{-\eta}}{(-q_\tau)}\left( \frac{\tau_{\rm i}^\eta-\tau_{\rm f}^\eta}{\eta}+
\Gamma(\eta,\tau_{\rm i})-\Gamma(\eta,\tau_{\rm f})\right)~~,\label{Snu-complete}
\end{equation}
where $\Gamma(\eta,\tau):=\int_\tau^\infty t^{\eta-1}e^{-t}dt$ is
the incomplete Gamma function.  Note that the frequency dependence is
in the Planck function and on the optical depth.  The reason for changing to
this formalism is that under the limit $\tau_{i}\rightarrow\infty$
and $\tau_{e}\rightarrow 0$ (valid for the frequencies where
we see  the spectrum is a power law), Equation \eqref{Snu-complete} simplifies to
\begin{equation}
S_\nu= 2 \times B_\nu(T_0) \Omega_0 \frac{\Gamma(\eta)}{q_\tau}(\tau_0(\nu))^{-\eta}~~, \label{Snu-simple}
\end{equation}
where $\Gamma(\eta)$ is the Gamma function evaluated for $\eta$. 
No assumption regarding the dependence of the opacity on frequency has been made. 
If we assume that $\tau_0(\nu)\propto\nu^{-2.1}$ --- a
valid approximation for the frequencies of interest --- we get the
Equation \eqref{eq-specindex} for the spectral index, equivalent to 
\[\alpha_{op}=2+2.1\eta~~,\] 
which is Equation (15) from \citet{Reynolds1986ApJ}. $\eta$ is a negative
number between $-1$ and 0. In particular, when  $\alpha_{op}=0.92$,
$\eta=-0.51$ and $\Gamma(\eta)<0$
$\left(\Gamma(-0.51)\approx-2\sqrt{\pi}\right)$. After introducing a
fiducial frequency $\nu_0$ and replacing it in 
Equation \eqref{Snu-simple}, we obtain Equation \eqref{eq-Snu}.  Equations
\eqref{Snu-complete} and \eqref{Snu-simple} 
{extend the derivation of} \citet{Reynolds1986ApJ}.

\subsection{Morphological constraints}
{Our} observations  give little information about the
physical scale size of the jet. Both ALMA  {and the centimeter wavelength observations} \citep{Guzman2010ApJ}
indicate that the jet is unresolved, which imply that the optically
thick portion  does not extend farther than the size of the
beam.  This condition is equivalent to 
\begin{align}
\bar{y}&=d \times \text{FWHM}_{\rm obs}/2\\
\tau(\bar{y})&=\tau_0 \left(\frac{\text{FWHM}_{\rm obs}\,d}{2 r_0 \sin i}\right)^{q_\tau}<1\qquad\text{(From Equation \ref{proj-height}.)}\label{size-cons}
\end{align}
where $\text{FWHM}_{\rm obs}$ is the full-width at half-maximum of the
synthesized beam.  It turns out that the most stringent constraint is
given by the ATCA data at 8.6 GHz.  All the models 
consistent with the data {easily fulfill} Equation \eqref{size-cons}.

{\section{Line flux expected from the jet model}\label{ap-hrls}}
To derive the expected HRLs, we start  in  Equation \eqref{Snu-intermedio}, 
which is also valid for the emission at the frequencies of the lines,
taking into account that the optical depth includes the continuum and line contributions. 
The equation for the total flux (continuum+recombination line) is
\begin{equation}
S_{\rm \nu, L+C} = 2\times B_\nu(T_0)\Omega_0 \int_{u_{\rm i}}^{u_{\rm f}} u^{\epsilon+q_T}%
\left(1-\exp\left(-\tau(\nu,u)\right)\right) du~~,\label{eq-SnuRL}
\end{equation}
where the integration is  over $u=y/y_0$ (see Figure \ref{fig-dibujo}). 
The total, continuum, and line optical depths are  given across the jet by
\begin{align}
\tau(\nu,u)&=\tau_C(\nu,u)+\tau_{L}(\nu,u)~~,\label{eq-Totalop}\\
\tau_{\rm C}(\nu,u) &= \tau_{\rm C,0}(\nu)u^{q_\tau}~~,\\
\tau_{\rm L}(\nu,u) &= \Tau_{\rm L}(u)\times\phi(\nu,u)~~.
\end{align}
 In the last equation, $\Tau_{\rm L}(u)$ represents the integrated
 optical depth of the line, and we assume that \mbox{$\int \phi(\nu,u)d\nu=1$}.
The line profile $\phi(\nu,u)$ depends on the line of sight, parameterized by $u$ in Equation \eqref{eq-SnuRL}.
  The integrated optical depth $\Tau_{\rm L}(u)$ depends on $u$
 because it depends on the density ($\propto N_e^2$) and on the path
 length.  We use the formulae for HRLs described in  \citet[\S\,2.3.5]{Gordon2002ASSL},
\begin{equation}
\Tau_{\rm L}=s(u)\times N_n \frac{\alpha h }{2 m_e} 
  f(n,\Delta n)\left(1-e^{-h\nu_{\rm L}/k T_e}\right)~~,\label{eq-Tau}
\end{equation}
where $s(u)$ is the path length, $N_n$ is the population in the $n$
quantum level, $\nu_{L}$ {corresponds} to the rest frequency of the line
associated with the $(n+\Delta n)\rightarrow n$ transition, and the rest
of the physical constants, including $\alpha\approx 137^{-1}$, are in
the usual notation.  The oscillator strength of the
line, $f(n,\Delta n)$, is  given by
\[f(n,\Delta n)=n\mathcal{M}(\Delta n)\left(1+1.5\frac{\Delta n}{n}\right)~~,\]
where $\mathcal{M}(1)=0.190775$ and $\mathcal{M}(2)=0.026332$ are
the Menzel constants \citep{Menzel1968Natur}.  The population level
$N_n$ under LTE is given by the Saha-Boltzmann ionization equation,
\begin{equation}
N_n=N_e^2\frac{n^2 h^3}{(2\pi m_e k T_e)^{3/2}}\exp\left(\frac{m_e c^2}{2 k T_e}\left(\frac{\alpha}{n}\right)^2\frac{m_H}{m_e+m_H}\right)~~,\label{eq-Nn}
\end{equation}
assuming a purely hydrogen gas with $m_H$ being the hydrogen
mass. Note that the statistical weight associated with degeneracy of the
$n$-th level, $g_n=n^2$, is already included.

As in the appendix for the continuum, we  assume that all
relevant quantities along a line of sight through the jet are given by
their values at the jet axis.

Note that the optical depth of the continuum and the integrated
optical depth $\Tau_{\rm L}$ are both proportional to $N_e^2$ and to
the path length $s$ (see Figure \ref{fig-dibujo}).  Therefore, the
quotient $\Tau_{\rm L}/\tau_{\rm C}$ is \emph{independent} of density
and path length and depends only on $T_e$. Under the assumption of
isothermality, it is also independent of the integration variable in
Equation \eqref{eq-SnuRL}.  We call this quotient the equivalent line width
of the transition, defined by $\Delta\nu_{\rm L}:=\Tau_{\rm
  L}/\tau_{\rm C}$. The value of $\Delta\nu_{\rm L}$ for the HRLs
observed in this work is given in Equation \eqref{eq-ew}.  We remark that
these are not truly line widths, but a measure of the area of the
line compared to the continuum level.

Then, we change the variable of integration to the  continuum opacity using
\begin{equation}
\begin{split}
\tau(\nu,u)&=\tau_{\rm C}(\nu,u)+\tau_{\rm L}(\nu,u)\\
&= \tau_{\rm C}(\nu,u)\left(1+\Delta\nu_{\rm L} \phi(\nu,u)\right)\\
&= \tau_{\rm C,0}(\nu)u^{q_\tau}\left(1+\Delta\nu_{\rm L} \phi(\nu,u)\right)~~.\label{eq-CoV2tau}
\end{split}
\end{equation}
This allows us to {calculate} the limit $\tau_{i}\rightarrow\infty$ and
$\tau_{e}\rightarrow 0$, which is the same procedure used to derive
Equation \eqref{eq-Snu}.
Therefore, Equation \eqref{eq-SnuRL} takes the following form,
\begin{equation}
S_{\rm\nu, L+C} =2\times B_\nu(T_0)\Omega_0\frac{(\tau_{\rm C,0})^{-\eta}}{(-q_\tau)} \int_0^\infty \tau^{\eta-1}\left(1-e^{-\tau (1+\Delta\nu_{\rm L}\phi(\nu,\tau))}\right) d\tau~~,
\end{equation}
where $\eta=(\alpha_{\rm op}-2)/2.1$.   Combining the previous equation
with Equation \eqref{eq-Snu}, we
{determine} that the line flux $S_{\rm L}=S_{\rm L+C}-S_{\rm C}$ is given
by
\begin{equation}
\begin{split}\label{Aeq-SL}
S_{\rm\nu, L}&=S_{\rm\nu, C}\left(\frac{\int_0^\infty \tau^{\eta-1}\left(1-e^{-\tau (1+\Delta\nu_{\rm L}\phi(\nu,\tau))}\right) d\tau} {-\Gamma(\eta)}-1\right)\\
&=S_{\rm\nu_0, C}\left(\frac{\nu}{\nu_0}\right)^{\alpha_{op}} \left(\frac{\int_0^\infty \tau^{\eta-1}\left(1-e^{-\tau (1+\Delta\nu_{\rm L}\phi(\nu,\tau))}\right) d\tau}{-\Gamma(\eta)}-1\right)~~,
\end{split}
\end{equation}
which gives the line flux density predicted for HRLs in LTE from the ionized jet model 
presented in the previous section.  Equation \eqref{Aeq-SL} corresponds to Equation \eqref{eq-SL}. 

%%%%%%%%%%%%%%%%%%%%%%%%%%%%%%%%%%%%%%%%%%%%%%%%%%%%%%%%%%%%%%%%%%%%%%
%% Bibliography
%%%%%%%%%%%%%%%%%%%%%%%%%%%%%%%%%%%%%%%%%%%%%%%%%%%%%%%%%%%%%%%%%%%%%%

\bibliographystyle{apj}
\bibliography{bibliografia}

%%%%%%%%%%%%%%%%%%%%%%%%%%%%%%%%%%%%%%%%%%%%%%%%%%%%%%%%%%%%%%%%%%%%%%
% Tables
%%%%%%%%%%%%%%%%%%%%%%%%%%%%%%%%%%%%%%%%%%%%%%%%%%%%%%%%%%%%%%%%%%%%%%
%%%%%%%%%%%%%%%%%%%%%%%%%%%
%% Continuum sources 

%%%%%%%%%%%%%%%%%%%%%%%%%%%%%%%%%%%%%%%%%%%%%%%%%%%%%%%%%%%%%%%%%%%%%%
%% Figures
%%%%%%%%%%%%%%%%%%%%%%%%%%%%%%%%%%%%%%%%%%%%%%%%%%%%%%%%%%%%%%%%%%%%%%

\end{document}